\documentclass[aps,pra,twocolumn,superscriptaddress,floatfix,a4]{revtex4}
\usepackage{epsfig,amsmath,amssymb,color}
\begin{document}

\title{BCS-BEC crossover and the disappearance of FFLO-correlations in a spin-imbalanced, one-dimensional Fermi gas}
\author{F. Heidrich-Meisner} 
\affiliation{Physics Department and Arnold Sommerfeld Center for Theoretical
Physics, Ludwig-Maximilians-Universit\"at M\"unchen, D-80333 M\"unchen,
Germany}
\affiliation{Institut f\"ur Theoretische Physik C, RWTH Aachen University, 52056 Aachen, Germany, and
JARA -- Fundamentals of Future Information Technology, Research Centre J\"ulich, 52425 J\"ulich, Germany}
\affiliation{Kavli Institute for Theoretical Physics, University of Santa Barbara, California 93106, USA} 
\author{A.E. Feiguin} 
\affiliation{Department of Physics and Astronomy, University of Wyoming, Laramie, WY 82071, USA}
\affiliation{Condensed Matter Theory Center, University of Maryland, College Park, MD 20742, USA}
\affiliation{Microsoft Project Q, University of California,  Santa Barbara,  California 93106, USA}
\author{U. Schollw\"ock} 
\affiliation{Physics Department and Arnold Sommerfeld Center for Theoretical 
Physics, Ludwig-Maximilians-Universit\"at M\"unchen, D-80333 M\"unchen, 
Germany}
\author{W. Zwerger} 
\affiliation{Physik Department, Technische Universit\"at M\"unchen, D-85747 Garching, 
Germany}

\date{February 26th, 2010}

%--------------------------------------------------------------------------
\begin{abstract}
We present a numerical study of the one-dimensional BCS-BEC crossover of a spin-imbalanced
Fermi gas. The crossover is described by the Bose-Fermi resonance model in a real space representation.
Our main interest is in the behavior of  the pair correlations, which, in the BCS limit, are of the 
Fulde-Ferrell-Larkin-Ovchinnikov type, while in the BEC limit, a  superfluid of diatomic molecules 
forms that exhibits quasi-condensation at zero momentum.
We use the density matrix renormalization group method to compute
the phase diagram as a function of the detuning of the molecular level and the polarization. As a main result,
we show that FFLO-like correlations disappear well below full polarization close to the resonance. The 
critical polarization depends on both the detuning and the filling.
\end{abstract}
%--------------------------------------------------------------------------

\maketitle
%**************************************************************************
% Introduction
%**************************************************************************

\section{Introduction}

Ultracold atoms provide a unique opportunity to study
basic many-body problems both in equilibrium and in
non-equilibrium situations \cite{bloch08}. A particularly appealing feature
of these systems is the possibility to change the interaction strength over 
a wide range via Feshbach resonances. In a two-component 
Fermi gas, this allows one to study the crossover from BCS-pairing
to a Bose-Einstein condensate (BEC) of strongly bound molecules
\cite{bloch08,ketterle08a,giorgini08}.
In a situation, in which the two states involved in the pairing are
equally populated, this is a smooth crossover. By contrast, in the case of an
imbalanced gas, unconventional superfluid ground states such as 
the Fulde-Ferrell \cite{fulde64} or Larkin-Ovchinnikov \cite{larkin64}
(FFLO) state with finite-momentum pairs, a Sarma phase with two Fermi surfaces \cite{sarma63}, 
or a mixture consisting of a BEC of strongly bound pairs 
and a Fermi gas of unpaired atoms have been proposed
\cite{son06,moon07,parish07,sheehy07,gubbels06}. 
 Experimentally, spin-imbalanced two-component Fermi gases have first
been realized at MIT \cite{zwierlein06,zwierlein06a,shin06} and Rice 
\cite{partridge06,partridge06a}. From the spin-resolved density profiles and,
in particular, the existence of a lattice of quantized vortices in a rotating gas \cite{zwierlein06a},
it is possible to observe the disappearance of a conventional superfluid in the center 
of the cloud with increasing imbalance.  Assuming that a local density approximation
applies, this allows one to determine the breakdown of BCS-type pairing beyond
a critical imbalance $p^{\mathrm{3D}}_c$ that is close to $p_{c}^{\mathrm{3D}}\sim 0.4$ for the uniform 
gas at unitarity in three dimensions  \cite{shin08,lobo06}.

Unfortunately, in the three dimensional (3D) case and in the unitary regime, where
the scattering length is much larger than the average interparticle
spacing, it is difficult, both experimentally and theoretically, to establish unambiguously
the existence of phases with unconventional pairing that are 
expected when the balanced ($p=0$) superfluid becomes unstable.
The experimentally observed density profiles \cite{shin08} at the unitary point 
are consistent with the prediction
of a first order transition from a balanced superfluid to a normal
state, in which the two spin components each form a Fermi liquid 
\cite{giorgini08}. This theoretical prediction is based on a variational ansatz
for the ground state \cite{lobo06,pilati08},  which excludes  
unconventional superfluid phases. It is therefore of 
considerable interest to study models, for which the phase 
diagram of the imbalanced gas along the BCS-BEC crossover 
is accessible by methods that are sensitive to states with complex order. 

In the case of  one dimension, such powerful numerical and analytical tools 
are indeed available. In fact, for both 
the attractive fermionic Hubbard model \cite{lieb68} and the 
associated continuum  model \cite{gaudin67,yang67}, there is an exact 
solution that can be extended
to the imbalanced case \cite{woynarovich91,orso07,hu07,zhao08,kakashvili09}.
The ground state phase diagram consists of three phases:
a balanced superfluid, a polarized intermediate phase and a fully
polarized, normal Fermi gas \cite{essler-book}. In the weak coupling limit,
both a solution of the Bogoliubov de Gennes equations \cite{machida84} 
and bosonization  \cite{yang01} indicate that the
polarized intermediate phase is an FFLO-like state at any
finite imbalance. This prediction has been recently verified  by
density matrix renormalization group (DMRG) \cite{feiguin07,tezuka08,rizzi08,luescher08,feiguin09a}  
and Quantum Monte Carlo (QMC) calculations \cite{batrouni08,casula08}.
It applies both to the continuum case and in the presence of an optical
lattice, and the FFLO state exists in mass-imbalanced systems as well \cite{batrouni09,wang09,burovski09,orso10}. Moreover, the one-dimensional (1D) FFLO state is also stable
in the inhomogeneous case that arises in the presence of 
a trapping potential \cite{feiguin07,tezuka08,casula08}. 
It is important to point
out that these methods give access to the regime of strong interactions as well, where the 
energy scale of the superfluid states is of the same order as the Fermi energy. In the context
of cold atoms, this is the relevant regime because in weak coupling, nontrivial order
only appears at unobservably low entropies of  $s\simeq T_c/T_F\ll 1$ per particle.     

 As realized by both Fuchs {\it et al.} 
\cite{fuchs04} and Tokatly \cite{tokatly04}, however, attractive
fermion models are not sufficient to account for the full physics of
the BCS-BEC crossover in one dimension.  Indeed, in the strong
coupling limit, they describe a Tonks-Girardeau gas of dimers.
They are unable, therefore, to cover the regime of weakly interacting 
bosons that is reached when the size of the two-particle bound state is 
smaller than the oscillator length of the transverse confinement. In this 
limit, the hardcore constraint of the tightly bound dimers 
becomes irrelevant.  
Moreover, in models of attractively interacting fermions 
there is only one phase at a finite spin imbalance below saturation, namely the FFLO phase  
\cite{yang01,orso07,hu07,feiguin07,batrouni08,tezuka08,zhao08, luescher08,rizzi08}.
As we shall emphasize in this work, the generic phase diagram of a
more general two-channel model is much richer, in particular, close to resonance.

A description of the 1D BCS-BEC crossover 
that properly accounts for the coexistence of fermions and bound 
pairs in the  imbalanced case can be achieved in the 
framework of the Bose-Fermi resonance model  
\cite{holland01,timmermans01} in which 
two fermions in an open channel couple resonantly to 
a diatomic molecule in a closed channel. The associated amplitude due
to the off-diagonal coupling between the open and closed channel 
determines the intrinsic width of the Feshbach-resonance 
\cite{bloch08}. 
In a continuum description, the 1D Bose-Fermi resonance model 
has been studied  by Recati {\it et al.} \cite{recati05}  for the 
special case of a vanishing imbalance, where a smooth 
BCS-BEC crossover occurs. Its BCS side is described by
attractively interacting fermions while on the BEC side, one 
has a repulsive Bose gas of dimers. In the limit of a broad Feshbach 
resonance, the transition between the two regimes is sharp, yet
continuous.  In particular, the quasi-long range superfluid order of
the ground state does not change along the full BCS-BEC crossover.
As realized recently by Baur {\it et al.}~\cite{baur09} in a study of the 
associated three-body problem, however, the situation is more complex
and interesting in the case of an imbalanced gas. There,  FFLO-physics
with spatially modulated pair correlations that  are  present  on the BCS-side
of the crossover  must disappear at a critical point, giving room to 
a Bose-Fermi mixture that is a conventional superfluid,
where quasi-condensation appears at zero total momentum.
 At the three-body level, this critical point shows up as a change in
the symmetry of the ground state wavefunction~\cite{baur09}. 

 As for studies on the many-body physics of the 1D Bose-Fermi resonance model, we refer the reader to  Refs.~\cite{recati05,orignac06,citro05,batchelor05,guan08}.
Bosonization  has been applied to the balanced case in Refs.~\cite{orignac06,citro05}, and Bethe ansatz results for the imbalanced case have been  presented in
Refs.~\cite{batchelor05,guan08}. 
FFLO correlations,  however,  have not been  discussed in 
either of these studies.

 Experimentally, the formation of molecules in Fermi gases that are
tightly confined in two transverse directions has been demonstrated 
by the ETH group \cite{moritz05}, using a balanced mixture. 
The binding energy of molecules is finite for an arbitrary sign of the 
3D scattering length $a$, in contrast to the situation without confinement,
where the two-particle binding energy vanishes on the BCS side of negative $a$. 

The objective of this work is to study a spin-imbalanced Fermi gas 
described by the Bose-Fermi resonance model Hamiltonian. We use a real-space representation
with a finite, incommensurate filling and map out the zero temperature phase diagram by computing  
pair correlations as a function of polarization and detuning. 
We find that FFLO correlations \cite{fulde64,larkin64} dominate in a wide parameter range,
and we clarify how the presence of  molecules affects the stability of this phase. 
Qualitatively, the presence of molecules binds
a certain fraction of minority fermions into molecules, reducing the overall number
of pairs in the FFLO channel. As a main result, we determine the critical polarization in the crossover region at which FFLO correlations disappear, and its dependence on filling and detuning.
Beyond this critical polarization and below saturation, the system is a superfluid of composite bosons  in the 
molecular channel immersed into a gas of either  fully or partially polarized fermions.
As a numerical tool, we employ the density matrix renormalization
group (DMRG) method \cite{white92b,white93,schollwoeck05}.

This exposition is organized as follows. First, in Sec.~\ref{sec:model}, we introduce the model Hamiltonian
and discuss its limiting cases. Further, in Sec.~\ref{sec:twobody}, we analytically solve the 
two-body problem. In Sec.~\ref{sec:results}, we present our DMRG results for the pair correlations, 
the momentum distribution,
and  the number of molecules as a function of filling, polarization and detuning. We close with 
a summary and discussion in Sec.~\ref{sec:sum}.

\section{The Bose-Fermi resonance model}
\label{sec:model}

\subsection{Hamiltonian}

We use a minimal Hamiltonian for the one-dimensional (1D) BCS-BEC crossover \cite{recati05,baur09}
in a real-space version, incorporating the kinetic energies of fermions and molecules,
the detuning of the molecular level, as well as the coupling between the
fermions and  molecules:  
\begin{eqnarray} 
H &=& - t \sum_{i=1}^{L-1}  (c_{i,\sigma}^{\dagger}   c_{i+1,\sigma} + h.c.) \nonumber \\
  &&  -t_{\mathrm{mol}} \sum_{i=1}^{L-1} (m_{i}^{\dagger}   m_{i+1} + h.c.) -(\nu+3t)\sum_{i=1}^L m_i^{\dagger}m_i\nonumber\\
  && + g \sum_{i=1}^{L} (  m_{i}^{\dagger} c_{i,\uparrow} c_{i,\downarrow} +h.c.) \label{eq:ham}\,.
\end{eqnarray}
$c_{i,\sigma}^{(\dagger)}$  is a fermionic annihilation (creation) operator acting on  site $i$,
while $m_{i}^{\dagger}$ creates a composite boson on site $i$.  The boson
energy is shifted with respect to that of single fermions by an effective detuning 
$\nu+3t$. It is chosen such that the energy for adding two fermions or one boson,
each at zero momentum, coincide at resonance $\nu=0$.     
The amplitude for the conversion of two fermions into a closed channel molecule 
and vice versa is given by the Feshbach coupling constant $g$. For a negative detuning
$\nu<0$ of the molecular level,  it gives rise to  an attractive two-particle interaction  $g^2/\nu<0$ between the fermions
 \cite{recati05}. 
Near resonance $\nu\simeq 0$,  this dominates any direct  background interaction $U_{\mathrm{bg}}$ between the two fermionic species,
which is therefore neglected from the outset. The hopping matrix elements for fermions 
and molecules are denoted by $t$ and $t_{\mathrm{mol}}$, respectively.
We further set $t_{\mathrm{mol}}=t/2$, which accounts for the mass ratio of 2$:$1 
between molecules and fermions. $L$ is the number
of sites. Further, $n_{i,\sigma}=c_{i,\sigma}^{\dagger}c_{i,\sigma}^{}$, yielding the number 
of fermions of each species as $N_{\sigma}=\sum_{i} \langle n_{i,\sigma} \rangle$,
with $N_f=N_{\uparrow}+N_{\downarrow}$ and the pseudo-spin index $\sigma=\uparrow,\downarrow$.
The only conserved particle number is $N=N_f + 2 N_{\mathrm{mol}}$, where $N_{\mathrm{mol}}=\sum_i \langle n_i^{\mathrm{mol}} \rangle$; $
n_i^{\mathrm{mol}}=m^{\dagger}_im_i$.
We use $n=N/L$   to denote the  filling factor and $p=(N_{\uparrow}-N_{\downarrow})/N$ as a 
measure of the polarization, which  we shall also sometimes refer to as imbalance. Note that at maximum one molecule can sit  on a single site, {\it i.e.}, 
the molecules behave as hard-core bosons.

\subsection{Two-body problem and spin gap}
\label{sec:twobody}

\subsubsection{Scattering amplitude and bound state energy }

In this section, we calculate the effective interaction between two 
fermions that is mediated by the molecules at the two-body level.
Following the method outlined in \cite{recati05}, the bound state
energy $\epsilon_b>0$ of two fermions is determined  by the condition
$$D_0^{-1}(k=0,\omega=-\epsilon_b)=\Pi(k=0,\omega=-\epsilon_b)\,,$$
where $D_0(k,\omega)$ 
is the bare molecular propagator and  
$\Pi(k,\omega)$ is the self-energy of the closed channel propagator (as usual, $\omega$ and $k$ denote frequency and momentum, respectively).
 
\begin{figure}[t]
\includegraphics[width=0.49\textwidth,angle=0]{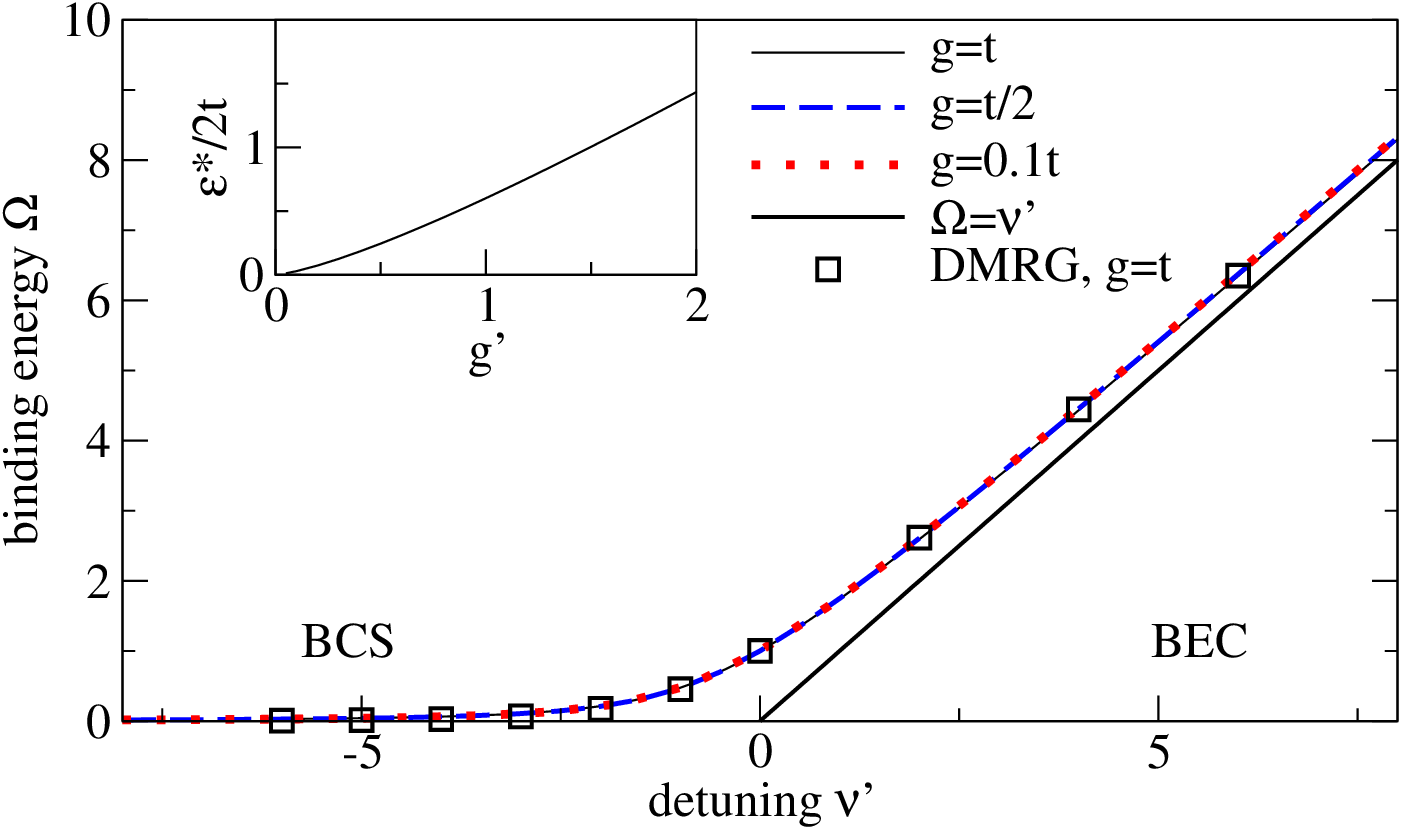}
\caption{(Color online) Dimensionless binding energy $\Omega$ vs. detuning as computed from Eq.~\eqref{eq:fomega}. The thick line is $\Omega={\nu'}$, the asymptotic behavior in the BEC regime (see the text). 
For comparison, the figure includes DMRG results (squares) for the spin gap $\Delta$ at $g=t$ and a small density of $n=0.1$, extrapolated in system size to the 
thermodynamic limit $L\to \infty$.
Inset: characteristic energy $\epsilon^*=\epsilon({\nu'}=0)$ vs $g'$.}
\label{fig:binding}
\end{figure}

The resulting equation
\begin{equation}
\epsilon_b -\nu =g^2\int_{-\pi}^{\pi}\frac{dk}{2\pi}
\frac{1}{\epsilon_b+4t(1-\cos{k})}
\label{bse}
\end{equation}
admits a unique, real solution $\epsilon_b>0$ irrespective of the sign of 
the detuning $\nu$. Of particular interest is the binding energy $\epsilon^{\star}=\epsilon_b(\nu =0)$
at resonance. Except for the scale $2t$ set by the bandwidth, it only depends  on the 
dimensionless Feshbach coupling constant $g'=g/(2t)$. For small coupling
strengths $g'\ll 1$, it is given by $\epsilon^{\star}/(2t)=g'^{4/3}/2^{2/3}$,
while $\epsilon^{\star}/(2t)=g'$ for $g'\gg 1$. The ratio 
$\epsilon^{\star}/(2t)=1/(r^{\star})^2$ is essentially the size of the bound state
(in units of the lattice spacing) at resonance. In terms of this characteristic length,
the condition for a broad Feshbach resonance is simply $nr^{\star}\ll 1$  \cite{recati05}.
Taking $\epsilon^{\star}(g')$
as a characteristic energy scale, the equation for the dimensionless
binding energy $\Omega=\epsilon_b/\epsilon^{\star}$ for an arbitrary value of the 
dimensionless detuning ${\nu'}=\nu/\epsilon^{\star}$ can be written in the form
\begin{eqnarray}
{\nu'}=-\sqrt{\frac{4+\epsilon^{\star}/(2t)}{\Omega \lbrack 4+\Omega\epsilon^{\star}/(2t)\rbrack}}+ \Omega\,,
\label{eq:fomega}
\end{eqnarray}
which is easily solvable for the bound state energy $\Omega({\nu'})$ as 
a function of the detuning. The definition of $\Omega$ guarantees that 
$\Omega\equiv 1$ at resonance, irrespective of the value of the Feshbach 
coupling $g'$. In Figure \ref{fig:binding}, we show the dependence of the binding energy 
$\Omega(\nu')$ on the detuning for three values of  $g/t=0.1,0.5,1$.  As suggested by the preceding
discussion, the $\Omega=\Omega(\nu')$-curve is practically independent of $g'$.

On the BCS side, where ${\nu'}\ll -1$, one obtains a very
small binding energy $\sqrt{\Omega}=\sqrt{4+\epsilon^{\star}/(2t)}/(2|{\nu'}|)\ll 1$,
approaching $\sqrt{\Omega}=1/|{\nu'}|$ for small values $g'\ll 1$
of the Feshbach coupling.  In the BEC regime of strongly
positive detuning ${\nu'}\gg 1$, the binding energy
$$\Omega={\nu'}+(g/\epsilon^{\star})^2/{\nu'}+\ldots$$ 
follows
the detuning, {\it i.e.}, the energy of the molecular state to leading order.
As a result, the closed channel fraction
\begin{eqnarray}
Z=\frac{\partial\epsilon_b}{\partial\nu}=1-\frac{(g/\epsilon^{\star})^2}{{\nu'}^2}+ \ldots
\label{Z}
\end{eqnarray}
is close to one, as expected in the BEC limit. The dimensionless binding
energy $\Omega=\left(r^{\star}/r_b\right)^2$ determines the size
$r_b$ of the bound state normalized to its value at resonance. 
For $\Omega\gg 1$, therefore, this size is much smaller than
the lattice spacing unless $g'\gg 1$.

\subsubsection{Spin gap}

 In the previous section, we argued that the binding energy $\Omega$ and in particular, $\epsilon^*$,
are important quantities to characterize the 1D BCS-BEC crossover on the two body level.
We next discuss the relation of $\Omega$ to the spin gap $\Delta$, which we calculate with 
DMRG as a function of filling, detuning, and the Feshbach coupling. The connection between the binding $\Omega$ 
and the spin gap has previously been pointed out by Orso \cite{orso07}.

 The spin gap is computed from 
 \begin{equation}
 \Delta(L)=E_0(S^z=1)-E_0(S^z=0)\,,
 \end{equation}
  where $E_0(S^z)$ is the ground-state energy of a system of length $L$ in the subspace with $S^z=(N_{\uparrow}-N_{\downarrow})/2$.
 We then extrapolate the finite-size data for $\Delta(L)$ in system size to the thermodynamic limit $L\to \infty$.
  
\begin{figure}[t]
\includegraphics[width=0.49\textwidth,angle=0]{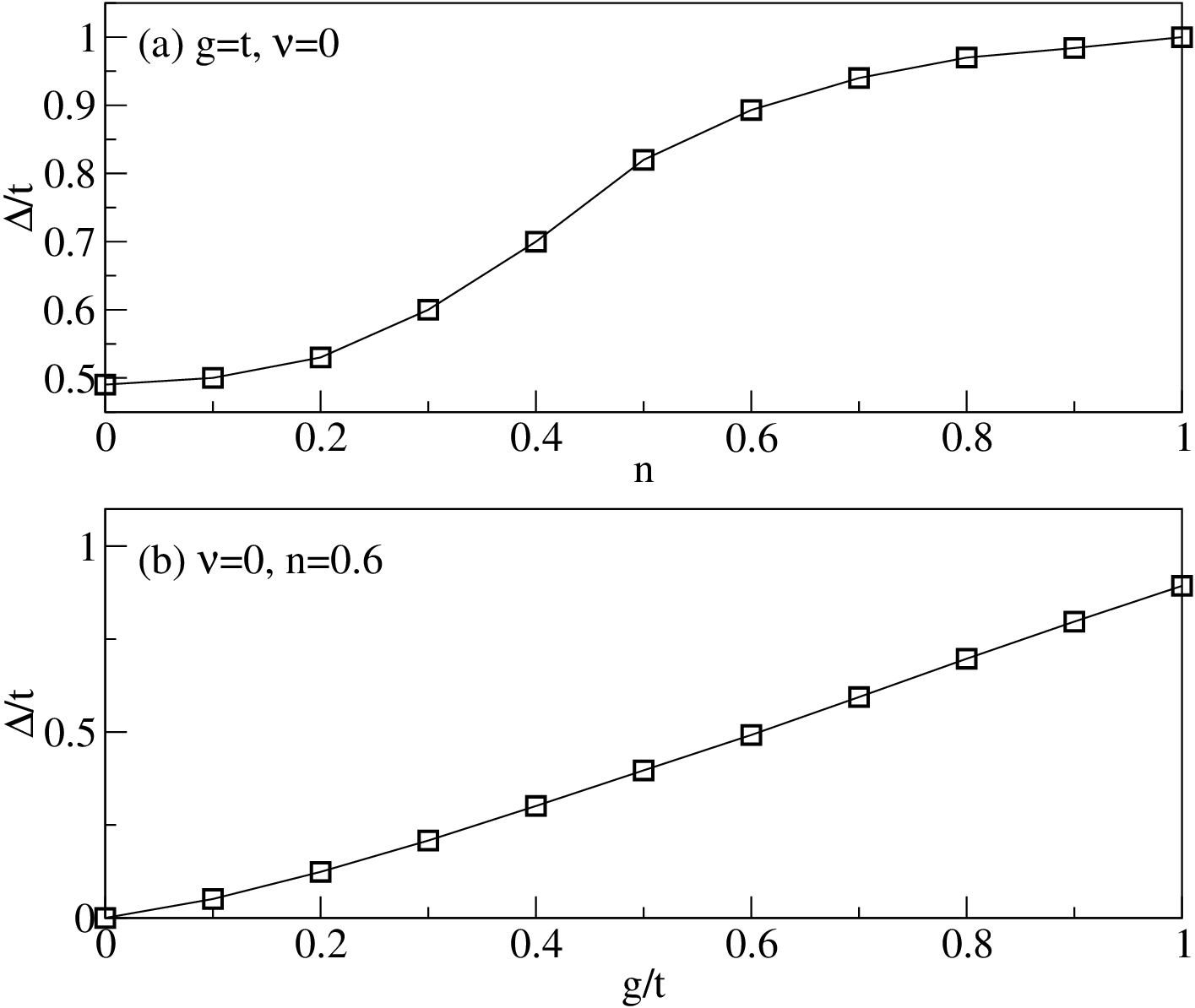}
\caption{(Color online) Spin gap $\Delta$ at resonance $\nu=0$ vs. (a) filling $n$ at $g=t$ and (b) vs. Feshbach coupling $g$ at $n=0.6$.
 All results are obtained by extrapolations in system size
to $L\to \infty$. } 
\label{fig:gap}
\end{figure}

 Figure~\ref{fig:binding} includes the DMRG data for the spin gap at a filling of $n=0.1$ 
and for $g=t$ (squares). 
Evidently, the spin gap coincides with the two-fermion binding energy $\Omega$ not only
on the BEC-side $\nu'>1$ where this is expected, but also far into the BCS regime.
Of course, for very weak coupling, this agreement must eventually be violated because
the spin gap $\Delta\simeq\exp{\lbrack-\pi/(2|\gamma|)\rbrack}$ depends on the filling $n$. 
In particular, it is exponentially small in the dimensionless coupling constant 
$|\gamma|=1/(2n|a_1|)\ll 1$ ($a_1$ is the effective scattering length in
one dimension, see~\cite{fuchs04}), while the two-particle binding 
energy $\epsilon_b=\epsilon^{\star}/{\nu'}^2$ is independent of $n$ and 
vanishes algebraically with the detuning in this regime.
Near resonance, the spin gap is identical with 
the two-particle binding energy in the low-density limit $nr^{\star}\ll 1$, 
as shown by Fuchs {\it et al.}~\cite{fuchs04}. 
With increasing values of the filling,
however, the spin gap increases, as is evident from Fig.~\ref{fig:gap}.
The many-body spin-gap is therefore clearly distinct from the two-particle binding energy. 

To illustrate this behavior, we  display $\Delta$ 
as a function of filling at $g=t$ in Fig.~\ref{fig:gap}(a) and as a function of $g$ at $n=0.6$ 
in Fig.~\ref{fig:gap}(b), both at resonance $\nu=0$.
$\Delta=\Delta(g)$ at $n=0.6$ also grows with the Feshbach coupling $g$.

\subsection{Limiting cases of the Bose-Fermi resonance model}
\label{sec:limits}
To guide the interpretation of our numerical results to be presented in the
following sections, we find it useful to start with a qualitative discussion of the limiting 
cases of the Hamiltonian Eq.~\eqref{eq:ham} in terms of the dimensionless detuning $\nu'=\nu/\epsilon^*$ 
[see also Ref.~\onlinecite{baur09}, which uses the more standard opposite sign 
convention for the detuning].

{\it (i) The BEC limit, $ \nu' \gg 1$ -- }  In this limit, all particles are bound in the molecular state,
{\it i.e.}, $N_{\mathrm{mol}}=N/2$. At  filling $N_{\mathrm{mol}}/L<1$, this realizes
a superfluid lattice gas of hardcore bosons, {\it i.e.}, effectively a Tonks-Girardeau gas of molecules.

Its ground state is characterized by quasi-long range order in the one-particle density matrix
\begin{equation}
\rho_{ij}^{\mathrm{mol}} = \langle  m_i^{\dagger} m_{j}\rangle\, \label{eq:opdm_mol}
\end{equation}
in the molecular channel of the form
$ |\rho_{ij}^{\mathrm{mol}} |\sim x^{-1/2}$ ($x=|i-j|$) \cite{mccoy68}.
 As the detuning 
is decreased and resonance is  approached, the molecules  start to make
virtual fluctuations into fermions. The presence of excess fermions suppresses
these fluctuations, giving rise to a repulsion between fermions and molecules
which is proportional to $g^2/\nu$ \cite{baur09}. Within a continuum model, 
this effective atom-molecule interaction on the BEC side of the resonance 
has  been calculated exactly at the three-body level 
by Mora {\it et al.} \cite{mora04}.  They find that 
the interaction is repulsive in the regime where the two-body binding energy 
$\epsilon_b$ is larger by a factor $2.2$ than its value 
$\epsilon^{\star}$ at resonance.  For smaller binding energies, on the BCS side, 
the effective atom-molecule interaction becomes attractive and also nonlocal, 
indicating that the picture of bosons that can coexist with unpaired fermions is 
no longer applicable  \cite{mora04,baur09}.   

 It is instructive to compare the regime $\nu'\gg 1$ of the lattice model studied here to the corresponding  continuum model studied in Ref.~\cite{recati05}. 
In the latter case, the relevant dimensionless interaction parameter $\gamma_B=g_B/n_B$  ($n_B$ denotes the density of molecules)
can be tuned to values small compared to one even in the deep 
molecular limit because
$g_B\sim|\epsilon_b|^{-5/2}$ vanishes as the two-particle binding energy 
$|\epsilon_b|$ becomes very large. As a result, the effective Luttinger exponent 
$K(\gamma_B)$ is then much larger than one and one obtains a weakly interacting
 gas of molecules, whose one-particle density matrix $\rho_{ij}^{\mathrm{mol}}$ decays as $|\rho_{ij}^{\mathrm{mol}}|\propto x^{-(1/2K)}$ with an exponent
$1/(2K)$ that is close to zero. In the continuum and for $\nu'\gg 1$, therefore,  
the weakly interacting molecule gas exhibits almost true long range order. 
This regime, however, is not reachable in the framework of the model 
Eq.~\eqref{eq:ham}, because even in the 
deep molecular limit ${\nu'}\gg 1$, where the size of the two-particle bound state
$r_b$ (in units of the lattice spacing, see the definition of $r_b$ given above) is much 
smaller than one, we still keep only the eigenvalues $0$ and $1$ for  the local
molecule occupation number $n_i^{\mathrm{mol}}=m^{\dagger}_im_i$. In
reality, however, more than one closed-channel molecule could sit on a lattice site in this
limit because the lattice spacing is much larger than $r_b$. We shall not further discuss
or pursue this question in the present work. 
Consequently, while we will be able 
to see the suppression of FFLO physics due to molecule formation, 
 which is the main focus of our present work,    
Eq.~\eqref{eq:ham} does not  
describe the full BCS-BEC crossover at a finite imbalance
that should  feature a weakly interacting BEC in the limit ${\nu'}\gg 1$. 

{\it (ii) The BCS limit, $\nu' \ll -1$ -- } Here, $N_{\mathrm{mol}}\approx 0$. Virtual
transitions into the molecular state give rise to a weak attractive on-site  interaction
$U=g^2/\nu$ between fermions. At a finite polarization $p>0$, we thus expect 
FFLO-like correlations with real-space oscillations in the modulus of the pair-pair correlations 
\begin{equation}
\rho_{ij}^{\mathrm{pair}} = \langle c^{\dagger}_{i,\uparrow} c^{\dagger}_{i,\downarrow} c^{}_{j,\uparrow}  c^{}_{j,\downarrow}
\rangle 
\,.\label{eq:rho_pair}
\end{equation}
For small polarizations, these correlations are described by the sine-Gordon
theory whose ground state is an array of domain walls, where the superfluid order 
parameter changes by $\pi$ \cite{machida84,yang01,hu07}. For 
larger polarizations, the domain walls merge and the order parameter 
acquires a purely sinusoidal form with a power law decay 
\begin{equation} 
|\rho_{ij}^{\mathrm{pair}} |\propto |\cos(Q x)|/x^{\alpha(p)}\label{eq:cos}
\end{equation}
as a function of the separation $|i-j|=x$. The associated wave vector 
\begin{equation} Q=\ k_{\mathrm{F},\uparrow}-\ k_{\mathrm{F},\downarrow}=\pi\, n\, p \,,\label{eq:Q} 
\end{equation}
is fixed by the density imbalance via the difference of  the Fermi-wave vectors
$ k_{\mathrm{F},\sigma}=\pi N_{\sigma}/L$ of the majority(minority) spins.
More precisely, as shown by Sachdev and Yang \cite{sachdev06}
from a generalized Luttinger theorem for Hamiltonians of the form \eqref{eq:ham}, 
the difference $ k_{\mathrm{F},\uparrow}- k_{\mathrm{F},\downarrow}$ of the Fermi wave vectors 
of the {\it interacting} system is quite generally fixed by the imbalance $p$ as
in Eq.~\eqref{eq:Q}. While the $N_{\sigma}$ are not conserved separately 
in the case where the bosons are condensed, this theorem implies
that the wave vector of superfluid order in the fermions is given by Eq.~\eqref{eq:Q},
independently of the detuning, {\it i.e.},  the strength of the interaction. 
In the notation of Ref.~\cite{parish07}, the associated FFLO state is thus
commensurate. 
 
The exponent $\alpha(p)$ of the power-law decay has a quite interesting
dependence on polarization and interaction strength, first discussed by 
Yang \cite{yang01}. At  vanishing polarization $p=0$, it is fixed by the 
Luttinger parameter $K_c>1$ of the attractive 1D Fermi gas in the charge sector 
via $\alpha(p=0) = 1/K_c$. In the limit of small polarizations, 
bosonization gives  $\alpha(p>0) = 1/K_c +1/2 $ \cite{yang01}, {\it i.e.}, 
a discontinuous jump of $\alpha(p)$ at $p=0^+$. This dependence 
has  recently been verified  in Ref.~\cite{luescher08}, using  
the attractive 1D Hubbard model.

\begin{figure}[t]
\includegraphics[width=0.49\textwidth,angle=0]{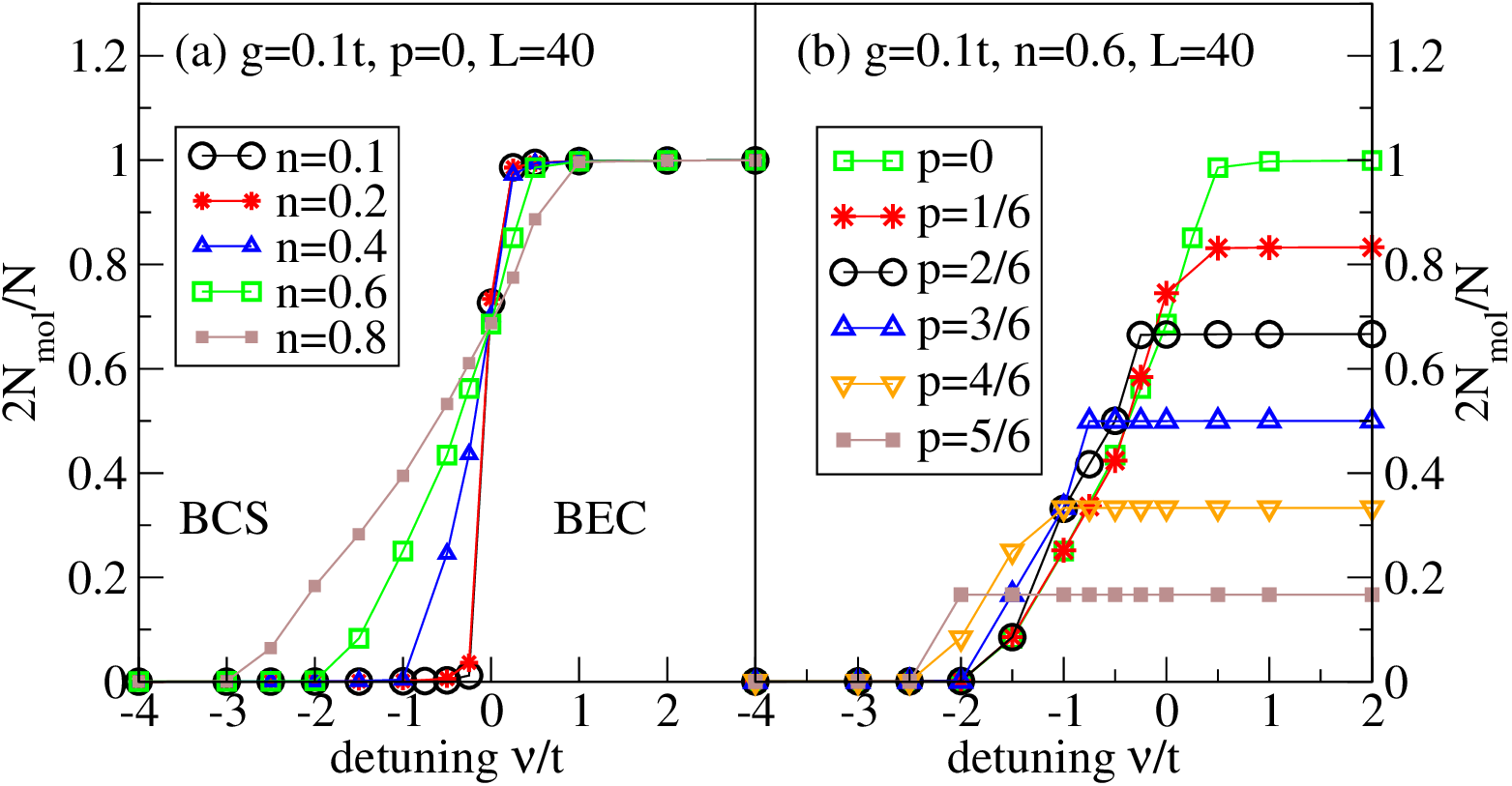}
\includegraphics[width=0.49\textwidth,angle=0]{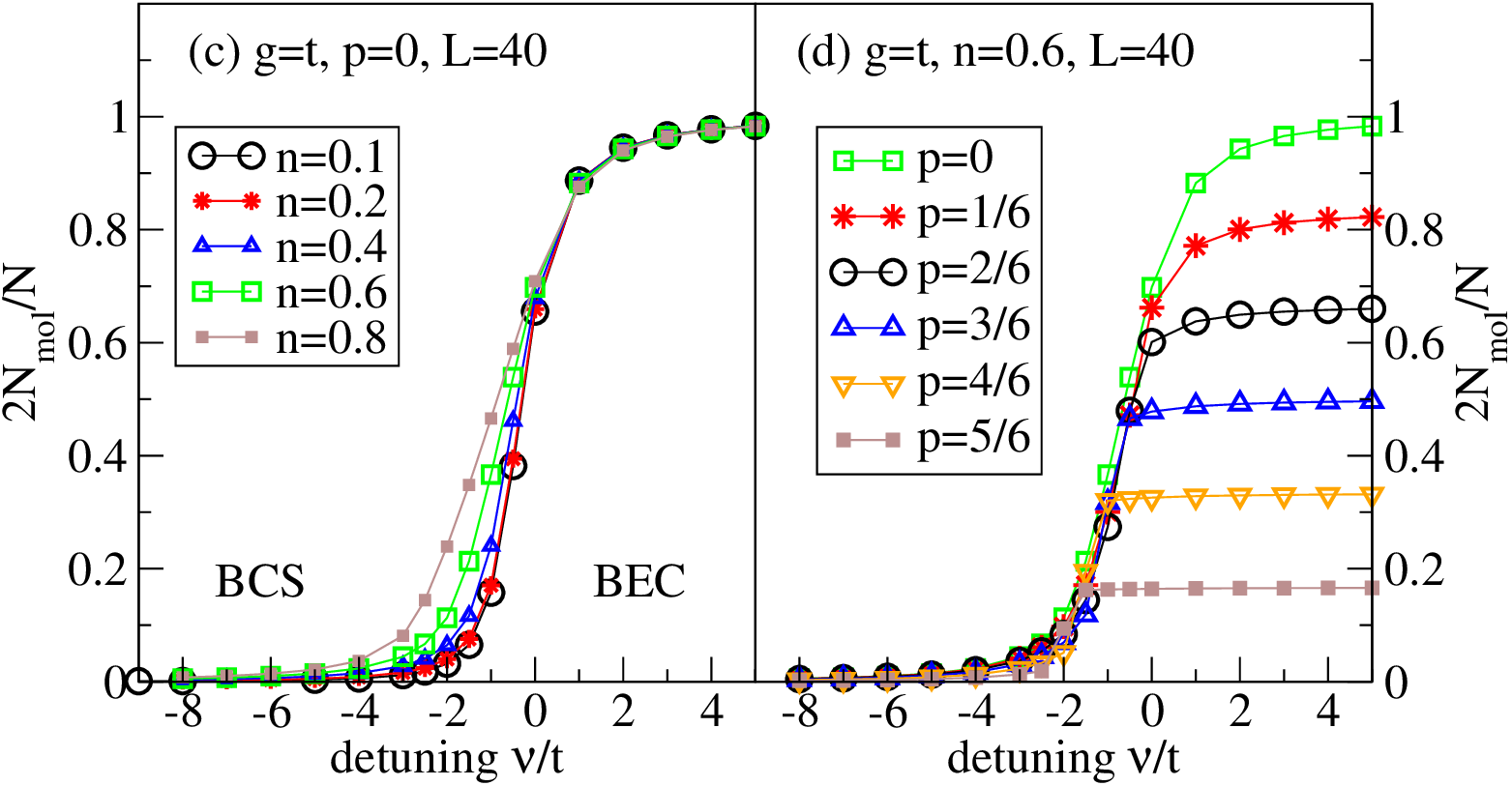}
\caption{(Color online) Number of molecules $N_{\mathrm{mol}}$ as a function of the detuning
$\nu$ for: (a),(b) a  resonance with $g=0.1t$ and (c),(d) a  resonance with $g=t$. 
(a) and (c): balanced mixture $p=0$, different fillings $n=0.1,0.2,0.4,0.6,0.8$ ($L=40$).
(b) and (d): Results for different polarizations
at a filling of $n=0.6$ ($L=40$). }
\label{fig:mol}
\end{figure}
  
\section{DMRG results for the imbalanced case}
\label{sec:results}

In this section, we present our DMRG results for the number of molecules, the 
pair correlations, the momentum distribution function (MDF) of both fermionic components, as
well as the MDF of the molecules, all as a function of polarization,
and detuning. 
As a main result we show that, while FFLO correlations are present in the BCS limit, 
as the number of molecules increases, the FFLO correlations disappear well below
full polarization. 
Upon increasing the polarization at a fixed detuning and in the crossover regime,
the system thus first has FFLO-like correlations, and then undergoes two phase transitions at polarizations $p_1$ and
$p_2$. For $p_1<p<p_2$, pairing at zero momentum 
 coexists with FFLO correlations,  
while for $p_2<p<1$, the system behaves as a Bose-Fermi mixture with only one fermionic component, the majority spins.
Therefore,  the large-$p$ phase is 
divided into a superfluid of molecules immersed into either a gas of partially polarized 
fermions or fully polarized fermions below saturation.
We further establish that the molecular and pair correlations are identical for $p<p_1$ in the sense that first, they
feature instabilities at the same wave vector and second, their highest occupied natural orbitals
are identical. Our results are summarized in  phase diagrams for $g=t/2$ and $g=t$ that are presented and  discussed in Sec.~\ref{sec:phase}.

\subsection{Number of molecules} 
\label{sec:mol}

To identify the crossover region characterized by a finite density of both fermions $N_f/L>0$ and molecules
$N_{\mathrm{mol}}/L>0$, we first calculate $N_{\mathrm{mol}}$ as a function of the detuning $\nu$ 
 at both $g=0.1t$  and $g=t$. The results are depicted in Fig.~\ref{fig:mol},
both for $p=0$ and several values of the filling $n$ [panels (a) and (c)] and $p>0$ at fixed filling $n=0.6$ [panels (b) and (d)]. 

We see that in the balanced  case, the crossover
region is between $-3t\lesssim \nu\lesssim t$ for $g=0.1t$ and in the range $-4t\lesssim \nu\lesssim 4t$ for $g=t$.
Moreover, the increase of $N_{\mathrm{mol}}$ as $\nu$ is moved from the BCS to the BEC side 
occurs over an increasingly wide range of detunings with increasing density $n$. This is 
consistent with the result that an abrupt change from a purely fermionic system ($N_{\mathrm{mol}}\approx 0$)
to a purely molecular one ($N_f\approx 0$) only exists in the low-density limit of a 
broad Feshbach resonance $nr^{\star}\ll 1$, as discussed 
previously in Refs.~\cite{fuchs04,recati05}.  An obvious, but important consequence 
of the off-diagonal Feshbach coupling $g$ is that the filling $n_f=N_f/L$ in the fermionic channel
depends on the detuning and the Feshbach coupling, ranging from $n_f=n$ in the $\nu'\ll -1$ limit
to $n_f=0$ in the BEC limit $\nu'\gg 1$. Therefore, the Fermi wave vectors $k_{F,\uparrow/\downarrow}$ vary, too.
This is consistent with our numerical observation from 
Fig.~\ref{fig:gap}, Sec.~\ref{sec:twobody}, that the spin gap is a function of $\nu$, $n$, and $g$.

The effect of the imbalance at some generic density $n$ [$n=0.6$ in Figs.~\ref{fig:mol}(b) and (d)]
is to make the window in which molecules and both  fermionic species coexist with comparable densities narrower. In the 
$g=0.1t$ case, the detuning, at which $2N_{\mathrm{mol}}\approx N$, is shifted towards the BCS regime $\nu<0$ as the polarization increases.
 
\begin{figure}[t]
\includegraphics[width=0.49\textwidth,angle=0]{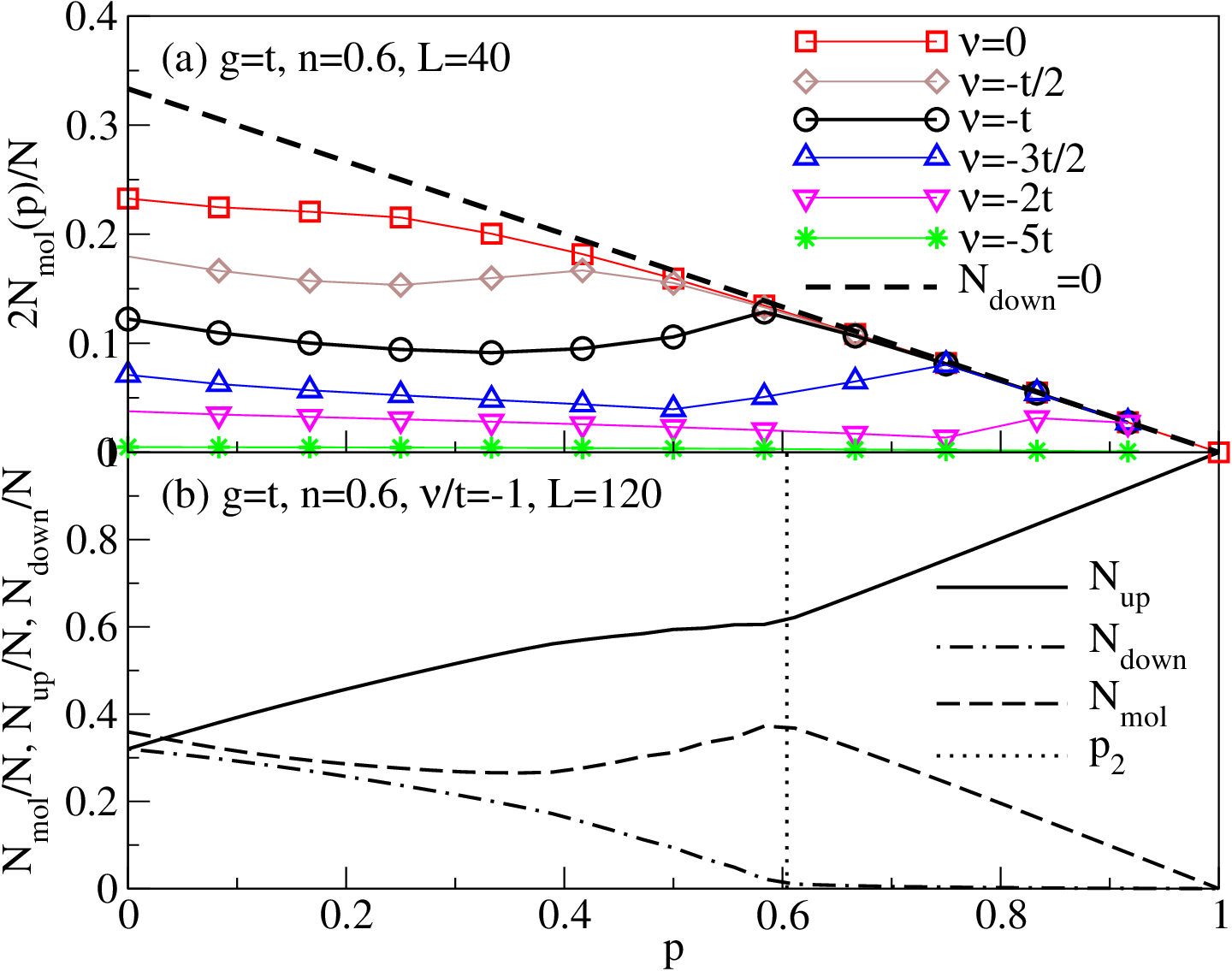}
\caption{(Color online) (a) Number of molecules $N_{\mathrm{mol}}$ as a function of polarization at $n=0.6$, $g=t$,
for several values of the detuning $\nu$. 
The dashed line shows the maximum possible $N_{\mathrm{mol}}$ at a given $p$ at which $N_{\downarrow}=0$.
The polarization $p_2$ at which $N_{\mathrm{mol}}(p,\nu)$ meets that line sets an upper limit for the
emergence of FFLO correlations. (b) Density of majority $N_{\uparrow}/N$ (solid lines), minority $N_{\downarrow}/N$ (dot-dashed lines), and molecules $2N_{\mathrm{mol}}/N$ (dashed lines) as a function of polarization 
for $\nu=-t$ ($n=0.6$, $L=120$, $g=t$). } 
\label{fig:nmol_g1}
\end{figure}

Figure \ref{fig:nmol_g1}(a) shows the number of molecules $2N_{\mathrm{mol}}/N$ as a function of polarization
and for several values of the detuning $\nu$ at $g=t$ and $n=0.6$. As soon as the line $N_{\downarrow}=0$
is reached at some polarization $p_2$, no pairing of fermions is possible anymore, and we are left with a BEC of molecules immersed into
a fully polarized gas of  fermions. This sets an upper limit, well below saturation $N=N_{\uparrow}$, for the emergence of FFLO-like
 correlations. In fact, in Sec.~\ref{sec:phase}, we shall see that  the FFLO regime 
 actually disappears well below $p_2$. 

It is further instructive to compare the polarization dependence of all particle densities, {\it  i.e.}, majority fermions $N_{\uparrow}/N$, minority fermions $N_{\downarrow}/N$, and molecules $N_{\mathrm{mol}}/N$, in the crossover region and before resonance $\nu= -t$, shown in Fig.~\ref{fig:nmol_g1}(b). The large-polarization
region, in which $N_{\downarrow}/N\approx 0$, is consequently characterized by a linear dependence of $N_{\mathrm{mol}}$ and $N_{\uparrow}$ on the polarization, with the slope
being independent of the detuning $\nu$. Note that from comparing
$L=40$ and $L=120$ sites data, we conclude that finite-size effects are negligible for the parameters considered. 

To determine $p_2$, we compute the polarization curves $p=p(h)$ for a given detuning and 
filling $n$, where $h$ denotes an effective 'magnetic field', coupled to the Hamiltonian through a Zeeman-like term
$$
H_{\mathrm{field}} =- h (N_{\uparrow}-N_{\mathrm{\downarrow}})
$$
that favors a finite imbalance $p>0$.

\begin{figure}[t]
\includegraphics[width=0.49\textwidth,angle=0]{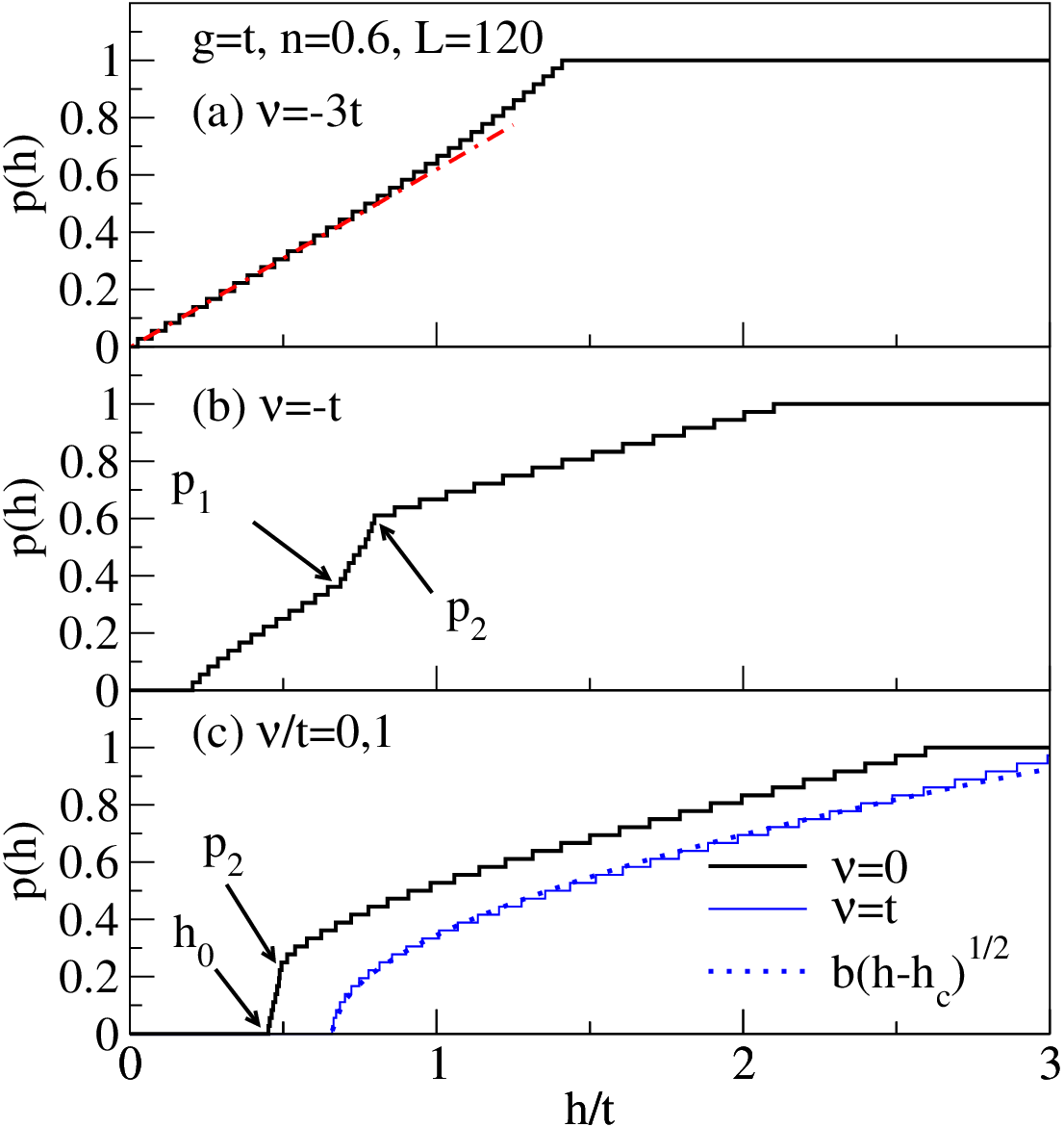}
\caption{(Color online) Polarization vs. field $h$ ($n=0.6$ and $g=t$) for: (a) $\nu =-3t$, (b) $\nu=-t$, and
(c) $\nu=0,t$ (thick and thin solid line, respectively). The dash-dotted line in (a) is a fit to $p(h)=a(h-h_c)$ close to $h_c$, while the dotted line in (c) is a fit of the numerical data  
to $p(h)=b \sqrt{h-h_{c}}$. $h_c$ is the critical field for the breakdown of the BEC at $p=0$ with $N\approx 2N_{\mathrm{mol}}$.} 
\label{fig:mag_curves}
\end{figure}

 The results for $g=t$ and $n=0.6$ are displayed for $\nu/t=-3,-1,0,1$ in Fig.~\ref{fig:mag_curves}. For $\nu=-3t$, the 
$p(h)$-curve has no features, and indicates the presence of a very small spin gap. At small polarization, $p=p(h)$ increases linearly with $h$, consistent with 
recent studies of the magnetization process of attractively interacting fermions \cite{vekua09,he09}.
At $\nu=-t$, we first identify the presence
of a large spin gap (identified by 2$h_c$), and two kink-like features  at finite polarizations $p_1$ and $p_2$. Essentially, at $p>0$, the system is a 
multi-component Luttinger liquid, and the presence of kinks indicates the disappearance or appearance of one component.
It is thus easy to guess that the kink at larger polarizations, {\it i.e.}, $p_2$ is associated with the depletion of the 
minority fermions, {\it i.e.}, $N_{\downarrow}\approx 0$ for $p>p_2$. This is consistent with our results for the particle densities
shown in Fig.~\ref{fig:nmol_g1}(b) and will be further corroborated by the discussion of the momentum distribution functions (see Sec.~\ref{sec:nmol}).
In view of the results for the BCS-BEC crossover of the imbalanced Fermi gas in 3D (see, {\it e.g.}, Refs.~\cite{gubbels06,sheehy07}), one might speculate about the possibility that phase
separation could appear also in one dimension. However, we stress  that the critical fields $h_1$ and $h_2$ corresponding to $p_1$ and $p_2$ are well separated. In particular,
a finite-size scaling analysis of the fields $h_1$ and $h_2$ for $\nu=-t$
shows that $h_2-h_1>0$ remains finite in the limit of $L\to \infty$. This rules out the possibility 
of a jump in $p(h)$ and thus of phase separation in a uniform system. 

The nature of the first kink $p_1$ in Fig.~\ref{fig:mag_curves}(b) will become obvious from the analysis of the pair correlations to be discussed in Sec.~\ref{sec:pairs}. As we shall see,
below $p_1$, we have pairs at a finite momentum ({\it i.e.}, the {1D FFLO} state), molecules and the two fermionic components, while at 
$p>p_1$, additional pairs at zero momentum are formed. 
On resonance, {\it i.e.}, at $\nu=0$, we still identify a kink at $p_2$,
while on the BEC side ($\nu=t)$, the polarization curve is smooth, 
 with  $p(h)\propto \sqrt{h-h_c}$, where the critical field $h_c$ for 
the onset of a finite polarization $p\ne 0$ is in fact connected to the spin gap
by the simple relation $2h_c=\Delta$ \cite{orso07}.

This behavior is characteristic for a band-filling transition of a single component, 
which in this case are the majority spins. Note that the same square-root
dependence in magnetization curves has been found for a 1D Bose-Fermi mixture \cite{guan08}.

\subsection{Pair correlations and superfluidity of molecules}
\label{sec:pairs}
\subsubsection{Momentum distribution functions for pairs, molecules, and fermions}
\label{sec:nmol}

To address the key questions of (i) the existence of FFLO-like correlations 
and (ii) their stability against the presence of molecules, we compute the 
momentum distribution function of first, pairs ($n_k^{\mathrm{pair}}$) 
and second, the momentum distribution
function  of the molecules ($n_k^{\mathrm{mol}}$) 
by taking a Fourier transformation of the real-space data 
for  Eq.~\eqref{eq:rho_pair} 
and of the one-particle density matrix of the molecules, $\rho_{ij}^{\mathrm{mol}}$ [compare Eq.~\eqref{eq:opdm_mol}],
respectively.
In the following we focus on $g=t$, unless otherwise stated.

\begin{figure}[t]
\includegraphics[width=0.49\textwidth,angle=0]{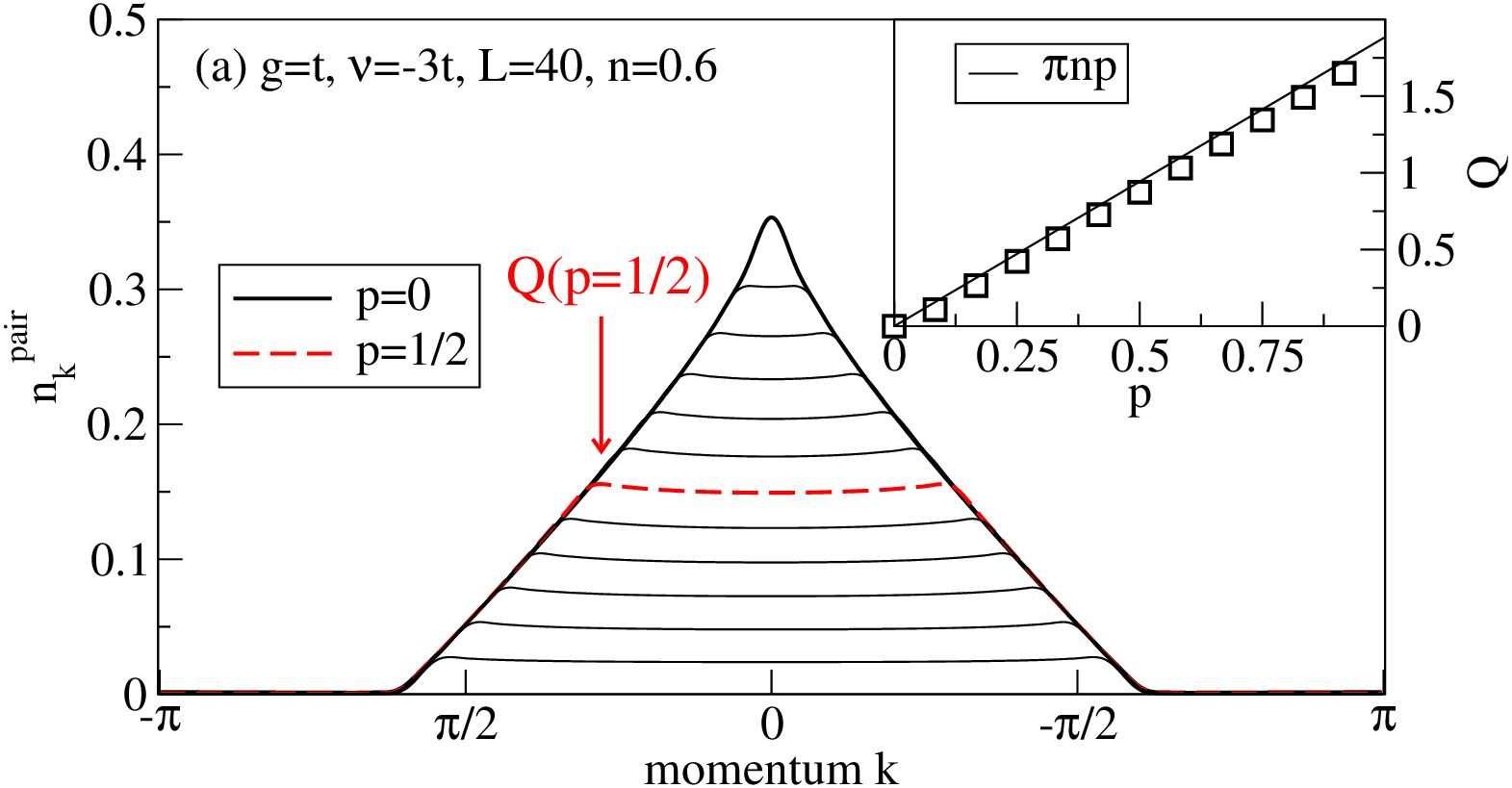}
\includegraphics[width=0.49\textwidth,angle=0]{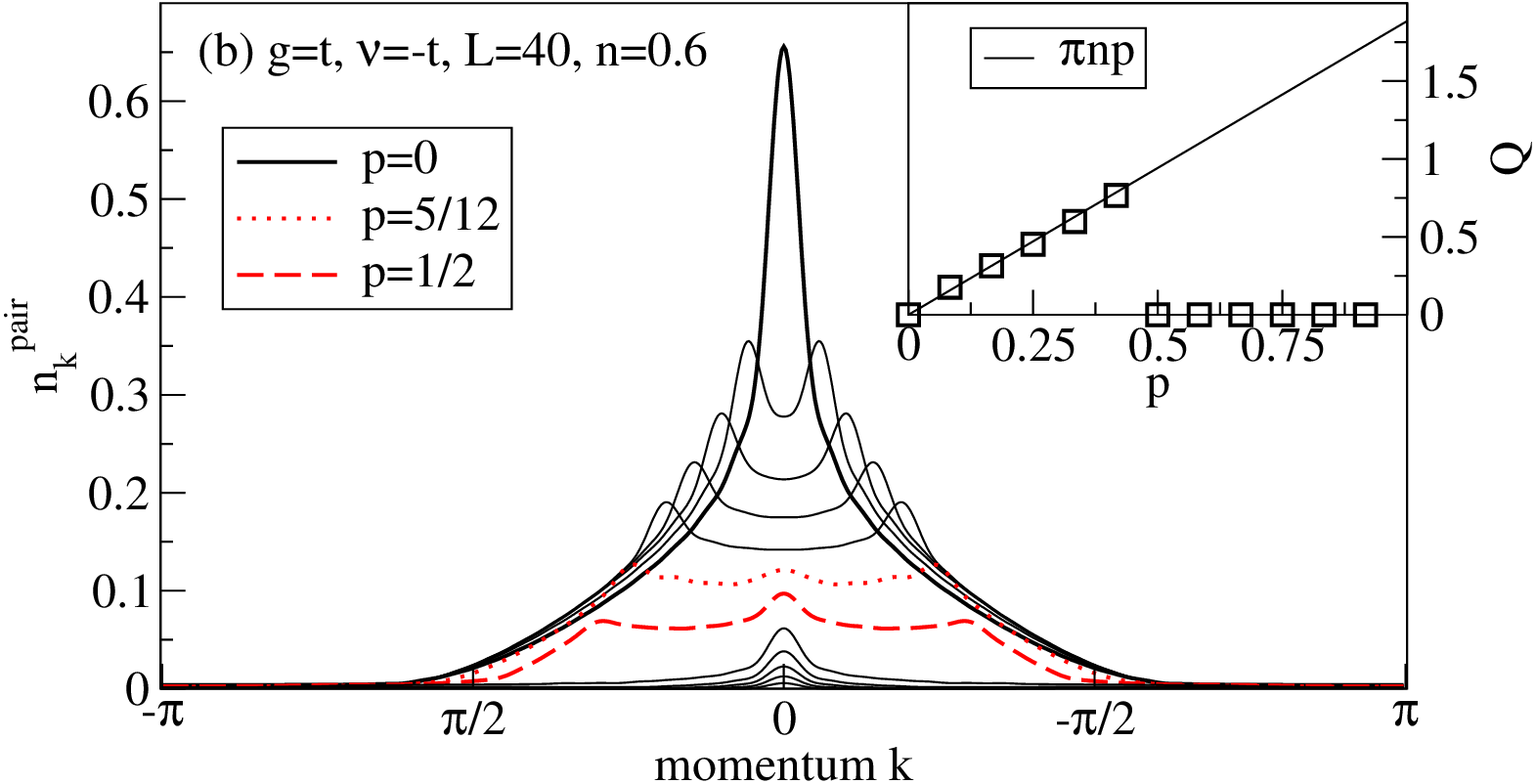}
\includegraphics[width=0.49\textwidth,angle=0]{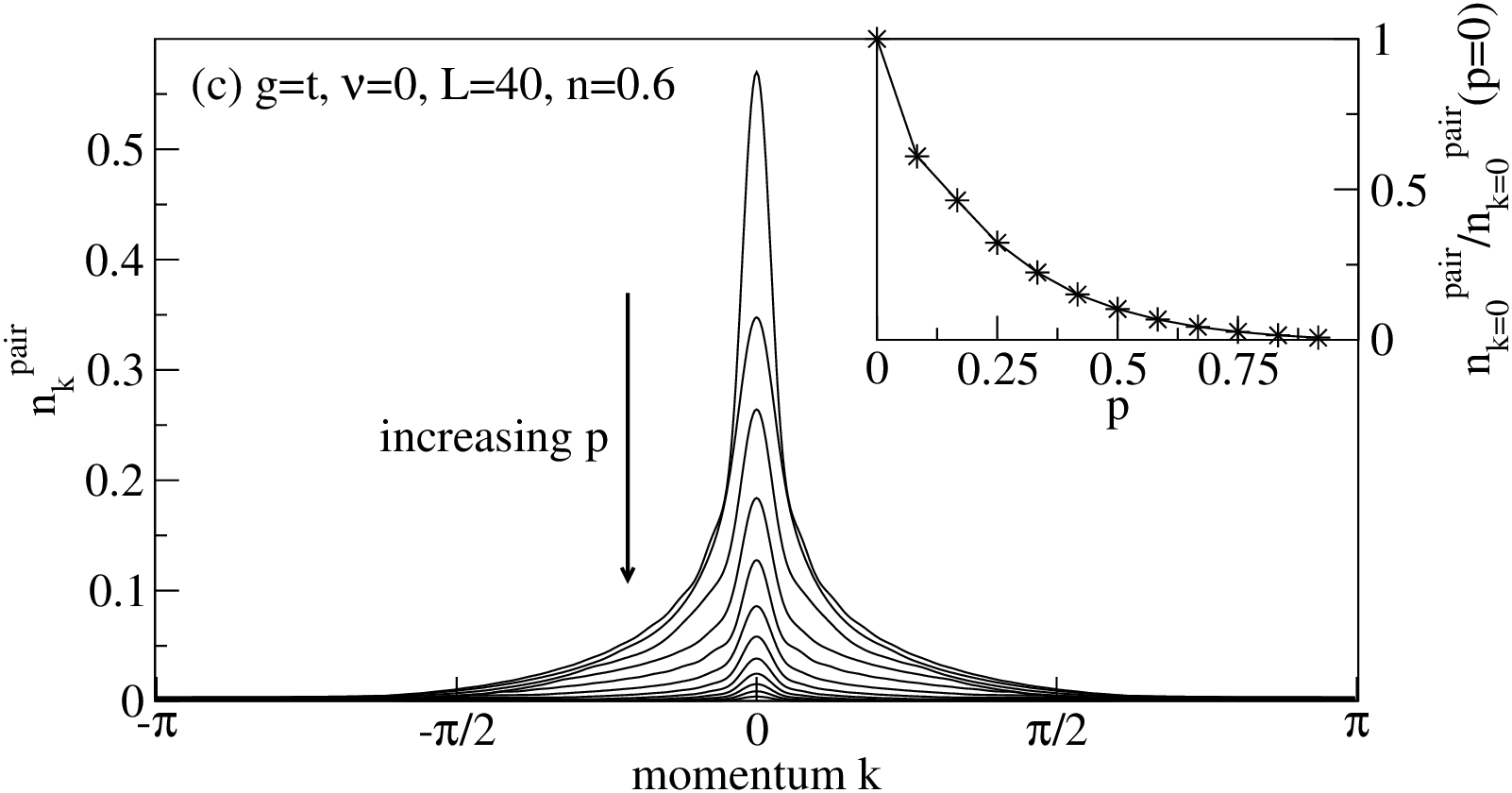}
\caption{(Color online) Fourier transform of pair correlations at $g=t$ and $n=0.6$, as a function of
polarization.  (a) $\nu =-3t$, BCS regime; (b) $\nu=-t$, crossover region; (c) $\nu=0$,
on resonance.
The insets in (a), (b) show the position $Q$ 
of the maximum in $n_k^{\mathrm{pair}}$ vs. polarization $p$ (squares).
The solid lines in these insets are $k_{F\uparrow}-k_{F\downarrow}=\pi n p$.  
 Inset in (c): $n_{k=0}^{\mathrm{pair}}$ vs. polarization.} 
\label{fig:nk}
\end{figure}

\begin{figure}[t]
\includegraphics[width=0.49\textwidth,angle=0]{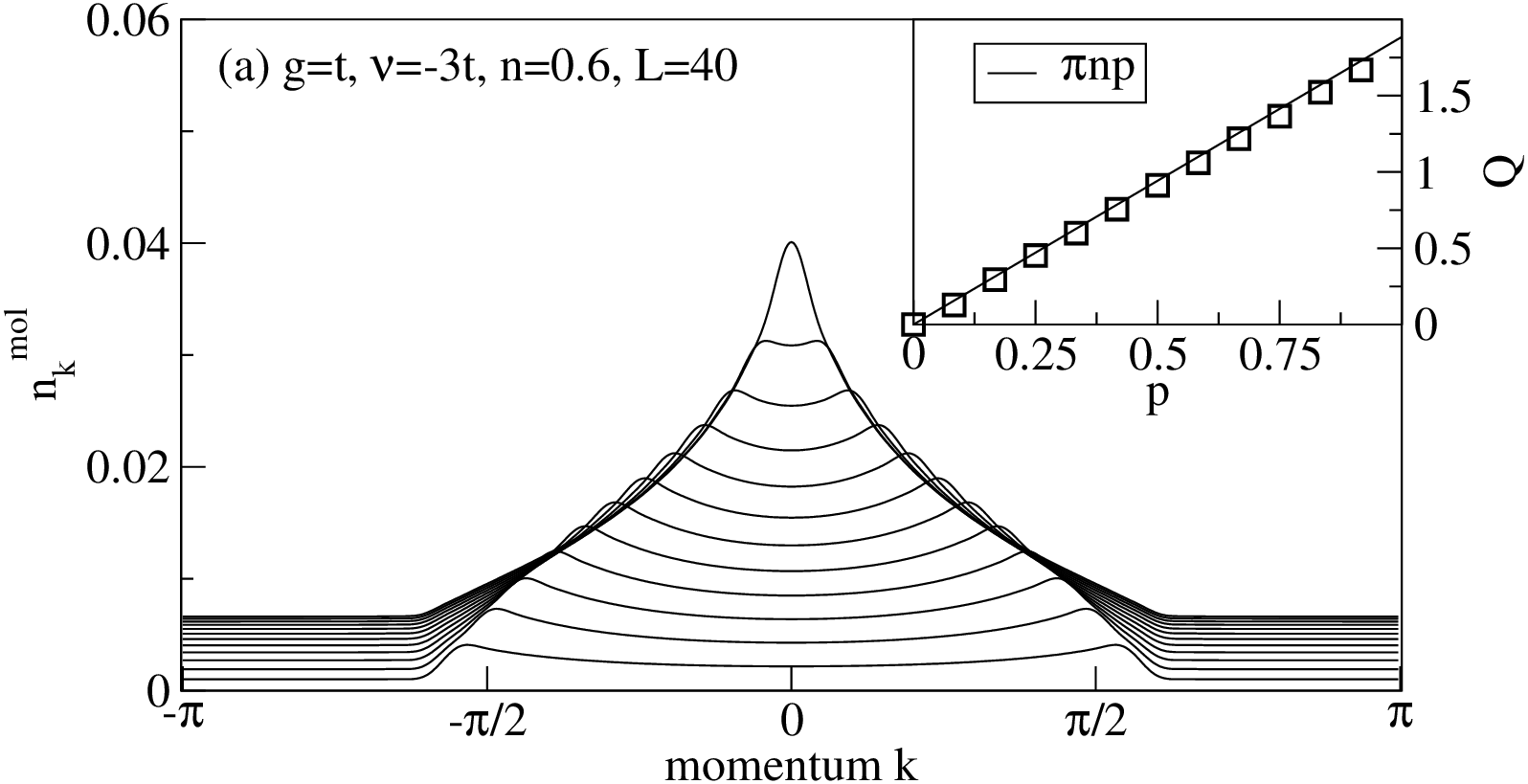}
\includegraphics[width=0.49\textwidth,angle=0]{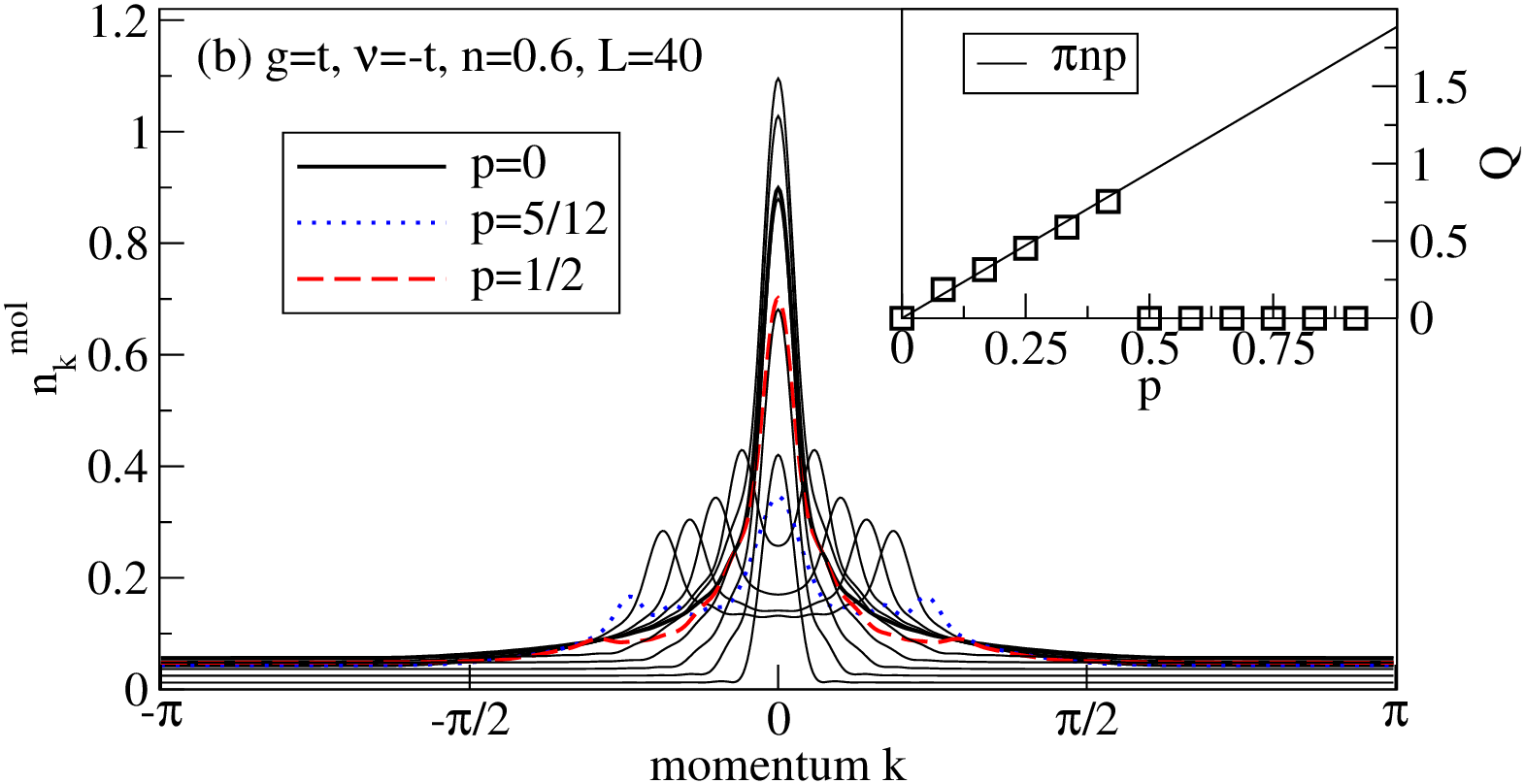}
\includegraphics[width=0.49\textwidth,angle=0]{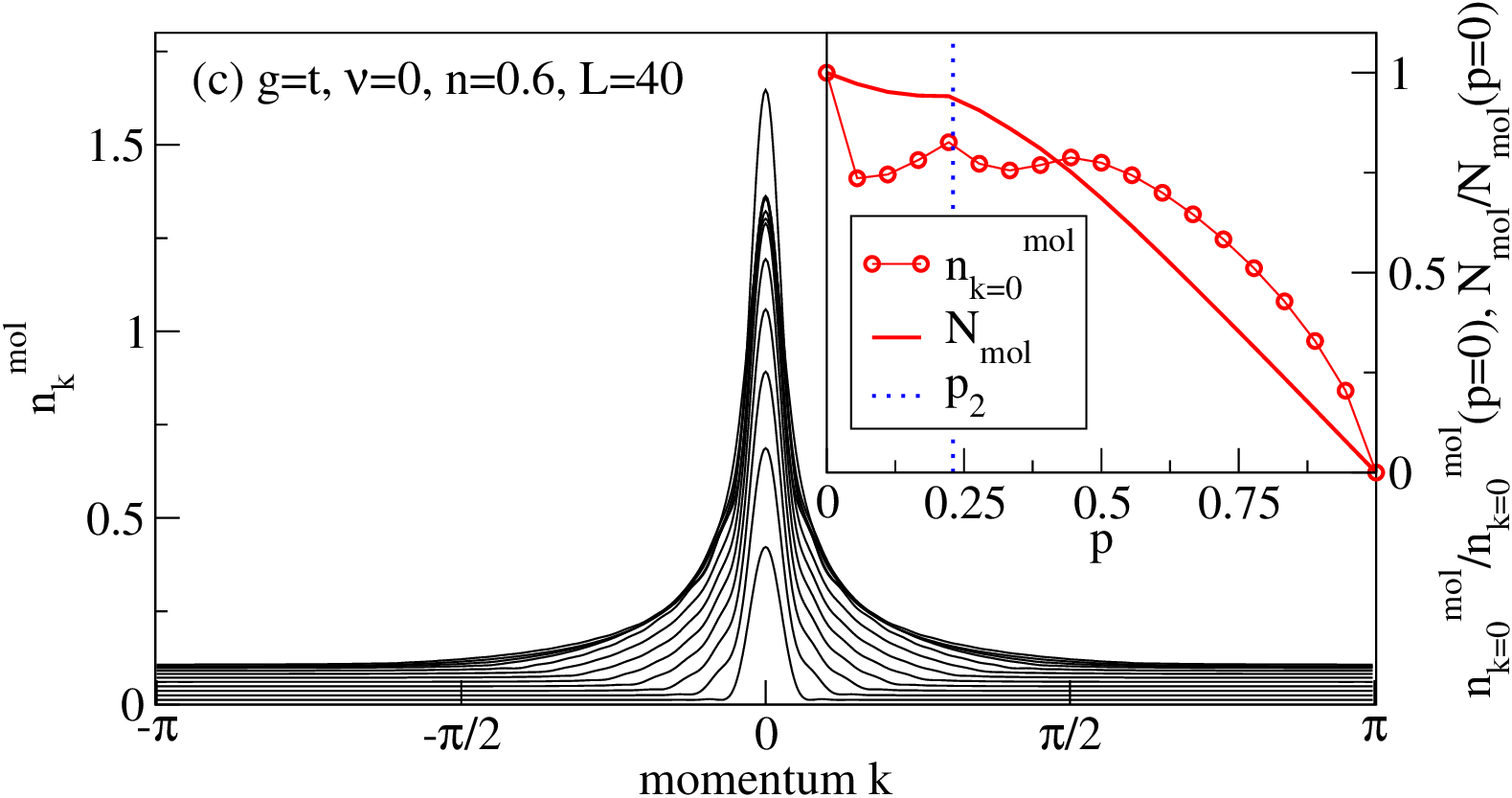}
\caption{(Color online) Momentum distribution of the molecules  at $g=t$ and $n=0.6$ as a function of
polarization.  (a) $\nu =-3t$, BCS regime; (b) $\nu=-t$, crossover region; (c) $\nu=0$,
on resonance.
The insets in (a) and (b) show the position $Q$ of the maximum in $n_k^{\mathrm{mol}}$ vs. polarization $p$. 
In the FFLO region, the MDF of the molecules shows a peak at $Q=k_{F,\uparrow}-k_{F,\downarrow}$.
 Inset in (c): $n_{k=0}^{\mathrm{mol}}/n_{k=0}^{\mathrm{mol}}(p=0)$ (circles) and  $N_{\mathrm{mol}}/N_{\mathrm{mol}}(p=0)$ vs. polarization ($L=60$).  }
\label{fig:nk_mol}
\end{figure}

\begin{figure}[t]
\includegraphics[width=0.49\textwidth,angle=0]{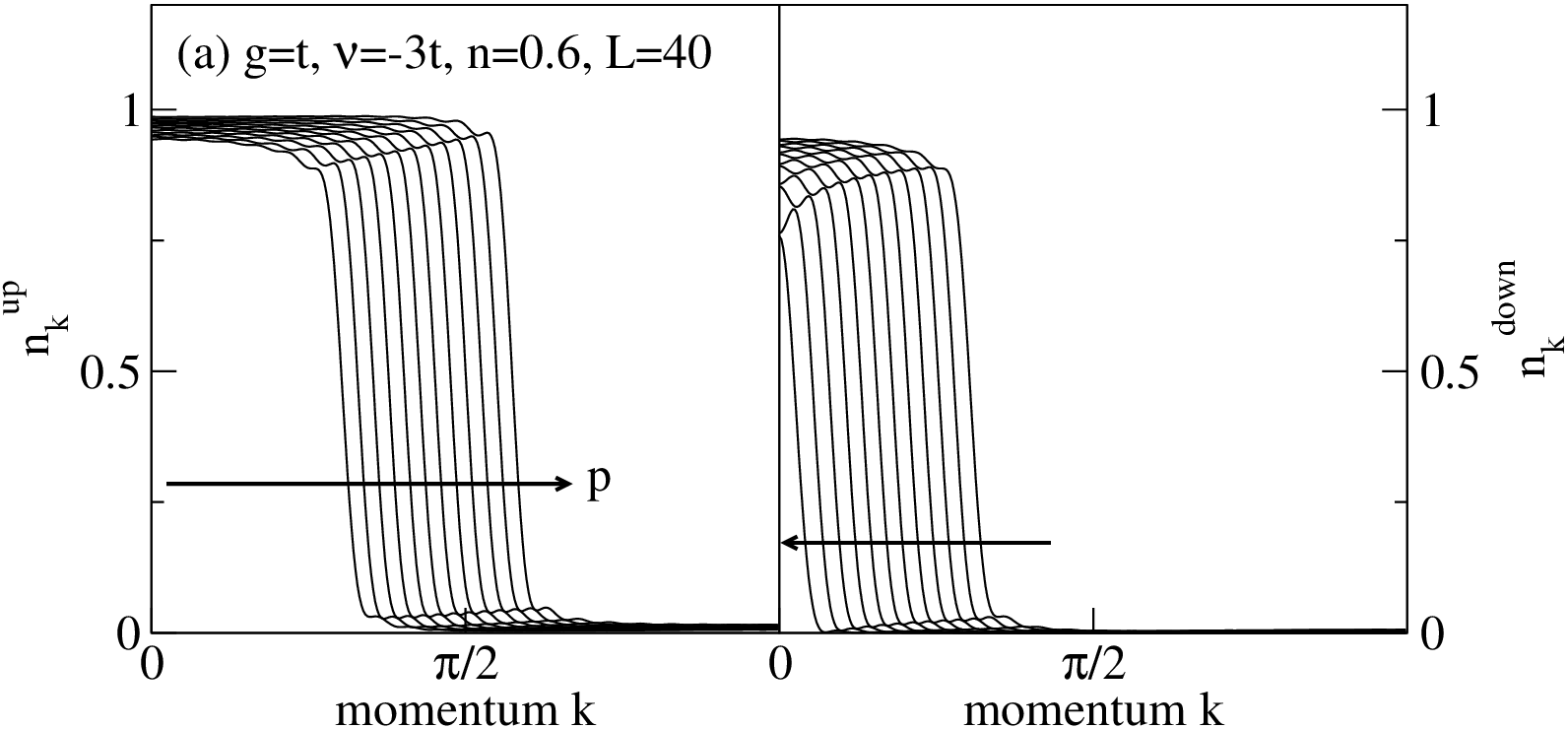}
\includegraphics[width=0.49\textwidth,angle=0]{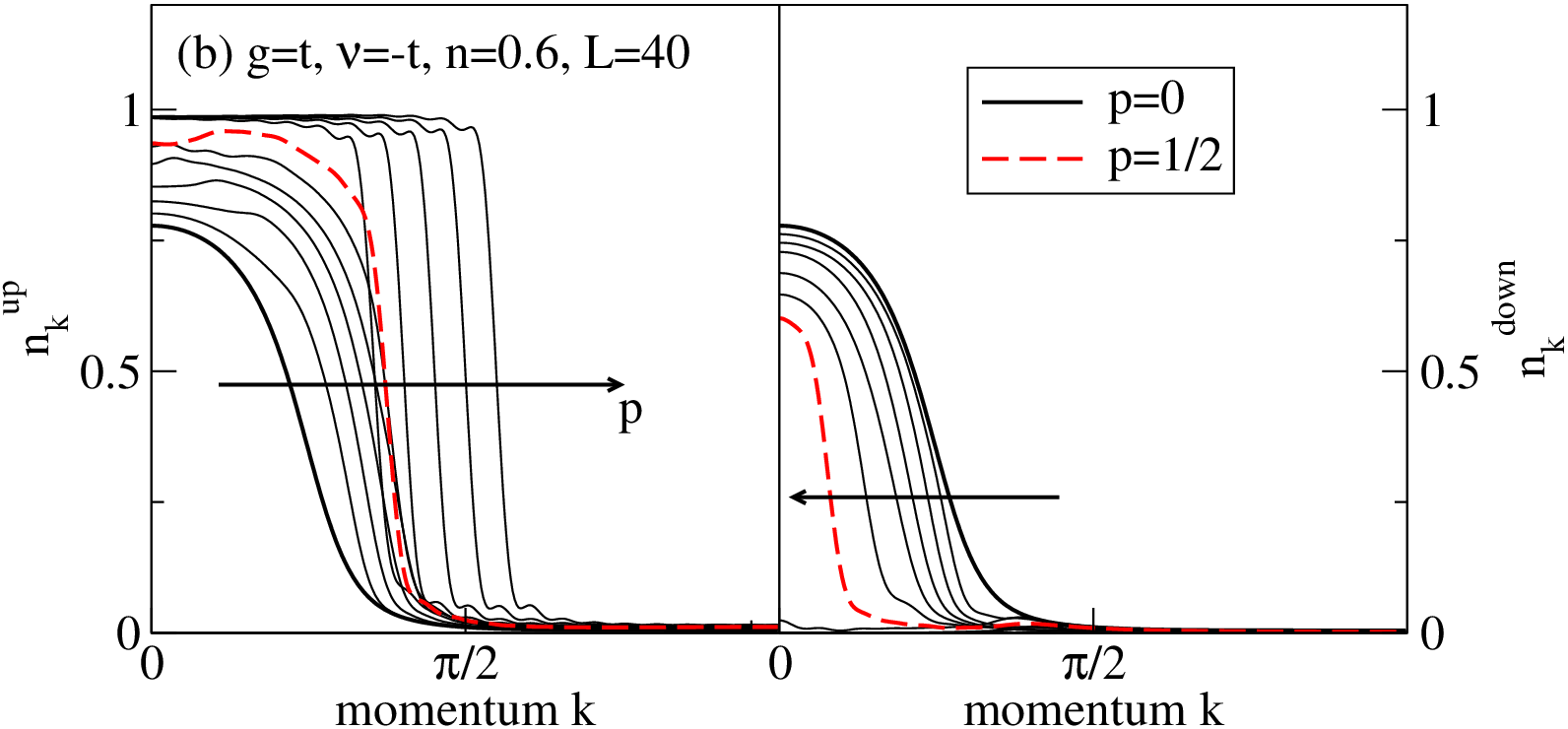}
\includegraphics[width=0.49\textwidth,angle=0]{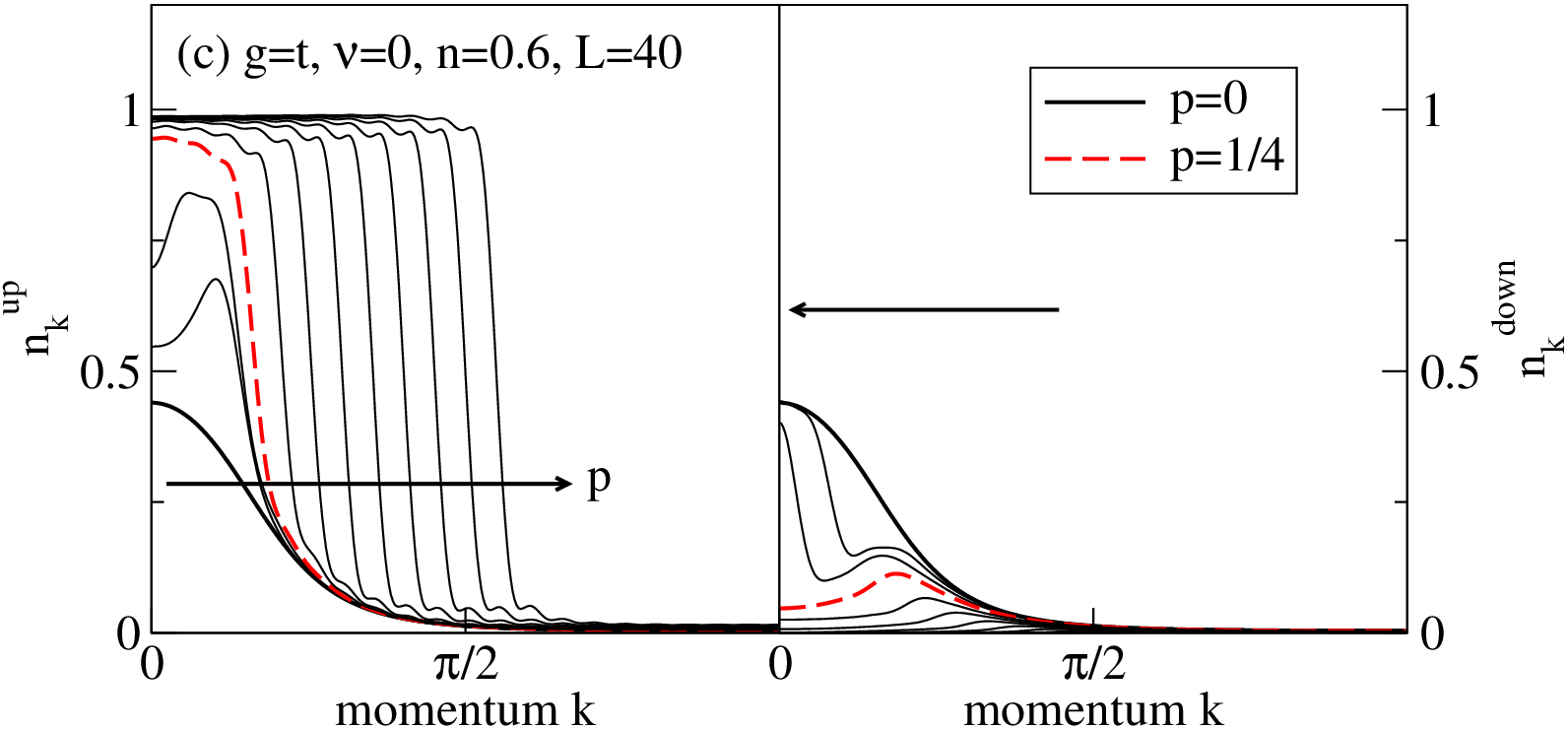}
\caption{(Color online) Momentum distribution functions $n_{k}^{\sigma}$ at $n=0.6$, $g=t$, as a function of
polarization [left panels $\sigma=\uparrow$; right panels: $\sigma=\downarrow$]. 
(a) $\nu =-3t$, BCS regime; (b) $\nu=-t$, crossover region; (c) $\nu=0$,
on resonance. Arrows indicate increasing polarization $p$.
 In the case of panel (b), $n_k^{\downarrow}\approx 0$ for $p>1/2$.
}
\label{fig:nk_mdf}
\end{figure}

The results for  $n_k^{\mathrm{pair}}$ and $n_k^{\mathrm{mol}}$ and a filling of $n=0.6$ are shown in Fig.~\ref{fig:nk} and Fig.~\ref{fig:nk_mol}, respectively.
 We choose three values of the detuning: $\nu=-3t$ [panels (a)], which is on the BCS side,
$\nu=-t$ [panels (b)] in the crossover region, and finally $\nu=0$ [panels (c)] on resonance.  
It is instructive to contrast the behavior of these  quantities with   that of the momentum distribution 
functions of majority and minority spins, {\it i.e.}, $n_k^{\uparrow,\downarrow}$, displayed in Fig.~\ref{fig:nk_mdf}.
 $n_k^{\sigma}$ is the Fourier transform of the one-particle density matrix $\rho_{ij}^{\sigma}=\langle
 c_{i,\sigma}^{\dagger}c_{j,\sigma}\rangle$.

Starting with the Fourier transform of pair correlations, we note that in the BCS limit and as the polarization is increased,
we observe quasi-coherence peaks at a finite momentum $Q>0$ [see Fig.~\ref{fig:nk}(a)]. Yet, these peaks are weak and the pairs' MDF resemble the one 
of a weakly interacting two-component Fermi gas described by the attractive Hubbard model [note that the finite-$Q$ peak is more pronounced in the molecules' MDF, Fig.~\ref{fig:nk_mol}(a)].  
 The rather weak peaks are probably a consequence of the fact that the pair correlations 
differ from a pure cosine [as suggested by Eq.~\eqref{eq:cos}]. This is certainly the case
at small values $p\ll 1$ of the polarization  
(see, {\it e.g.}, Ref.~\cite{machida84} and the discussion in Sec.~\ref{sec:limits}).

The position $Q$
of the maximum in $n_k^{\mathrm{pair}}$ follows $k_{F,\uparrow}-k_{F, \downarrow}$, as we illustrate in the insets of panels (a) and (b) in Fig.~\ref{fig:nk}.
This, as usual, is a defining feature of the {1D FFLO} state.

The quasi-coherence peaks are way more pronounced in the crossover region, {\it i.e.}, $\nu=-t$, which is of 
primary interest in this work [see Fig.~\ref{fig:nk}(b)]. We observe the 
breakdown of FFLO-like correlations at a finite polarization $0<p_c^{\mathrm{1D}}<1$. This
critical polarization $p_c^{\mathrm{1D}}$ is smaller than the upper limit $p_2$ discussed
above. An emergent feature of the pairs' MDF in the crossover region $\nu\sim -t$ 
 is  the 
coexistence of peaks at both $Q=0$ and $Q>0$ at intermediate polarization [see, {\it e.g.}, the dotted line in Fig.~\ref{fig:nk}(b)].
By determining the polarization at which we see pairing at both $Q=0$ and $Q>0$ [the dotted line in Fig.~\ref{fig:nk}(b)],
we find that this coincides with the first kink seen at $p_1$ in the polarization vs. magnetic field curves shown in Fig.~\ref{fig:mag_curves}(b).
Therefore, we conclude that the first phase transition and thus the boundary of the {1D FFLO} phase in the crossover regime and at $p>0$ is the one at $p=p_1$ where  pairing at $Q=0$
starts to contribute, effectively adding  an additional quasi long-range order parameter 
to the system. 
We can further define a crossover polarization $p^*>p_1$, beyond which the dominant instability 
is at $Q=0$. 
In the example of $\nu=-t$ shown in Fig.~\ref{fig:nk}(b), $p^*=1/2$.
Note that slightly above $p^*$, some modulation
in the pairs' MDF survives, which shows up as a smaller maximum in $n_k^{\mathrm{pair}}$ at a finite momentum. 
Finally, we note that the FFLO correlations are typically enhanced at low densities ({\it e.g.}, at $n=0.2$; results not
shown here). To summarize, we identify $p_c^{\mathrm{1D}}$ with the upper boundary of the FFLO phase, {\it i.e.}, $p_c^{\mathrm{1D}}=p_1$.

Right at resonance ($\nu=0$), no signatures of FFLO correlations are visible any more, and the momentum distribution functions of both the pairs and the molecules 
feature a maximum at zero momentum
[see Figs.~\ref{fig:nk}(c) and \ref{fig:nk_mol}(c)]. We observe the same behavior on the BEC side, $\nu >0$.
For illustration, the $k=0$-weight  in the pair and molecular MDFs
are shown as a function of polarization in the insets of Figs.~\ref{fig:nk}(c) and \ref{fig:nk_mol}(c).  
Quite notably, $n_{k=0}^{\mathrm{mol}}$ exhibits features that can be related to the phase transitions the system undergoes as $p$ increases.
First, the weight discontinuously drops from its $p=0$ value, as the critical field for breaking up molecules is overcome at $p=0^+$.
Second,    $n_{k=0}^{\mathrm{mol}}$ takes a maximum at $p_2$, where the system enters into the Bose-Fermi mixture phase at $p>p_2$.
A similar, yet less significant behavior can be seen in the number of molecules, $N_{\mathrm{mol}}(p)/N_{\mathrm{mol}}(p=0)$, which we have included
in the inset of Fig.~\ref{fig:nk_mol}(c) for comparison (solid line) [see also Fig.~\ref{fig:nmol_g1}(b)].

 An important point that should be emphasized in this context is the fact that 
the respective quasi-condensates of molecules and fermions are locked into 
each other. Indeed, they qualitatively show the same behavior concerning the position of their maxima,
as is evident from comparing Figs.~\ref{fig:nk} and \ref{fig:nk_mol}.

We next discuss the MDF of the two fermion components, shown in Fig.~\ref{fig:nk_mdf}.
In the BCS limit, the MDFs feature a sharp edge, reminiscent of a weakly interacting lattice gas and consistent with the features observed in Fig.~\ref{fig:nk}(a). 
As $\nu$ moves the system
into the BEC regime, the $p=0$ MDFs become quite broad, as expected for 
a strongly interacting system and and for the standard BCS-BEC crossover (see, {\it e.g.}, Refs.~\cite{giamarchi,bloch08}). 
Upon polarizing the system, $n_{k}^{\uparrow}$ develops a sharper edge [see Figs.~\ref{fig:nk_mdf}(a),(b), and (c), left panels], as eventually, only
the majority fermions remain. This is particularly evident in the case of $\nu =-t$ shown in Fig.~\ref{fig:nk_mdf}(b): for $p> 1/2$,
$N_{\downarrow}=\sum_k n_k^{\downarrow}\approx 0$. Simultaneously, for $p>1/2$,  $n_{k}^{\uparrow}$ changes from a smooth function seen at $p\leq 1/2$ 
to a steep one, since for $p>1/2$, there is a single fermionic component left.
Thus the depletion of minority fermions characterizes the transition to the Bose-Fermi mixture phase at $p\geq p_2$.

\subsubsection{Natural orbitals}

\begin{figure}[t]
\includegraphics[width=0.49\textwidth,angle=0]{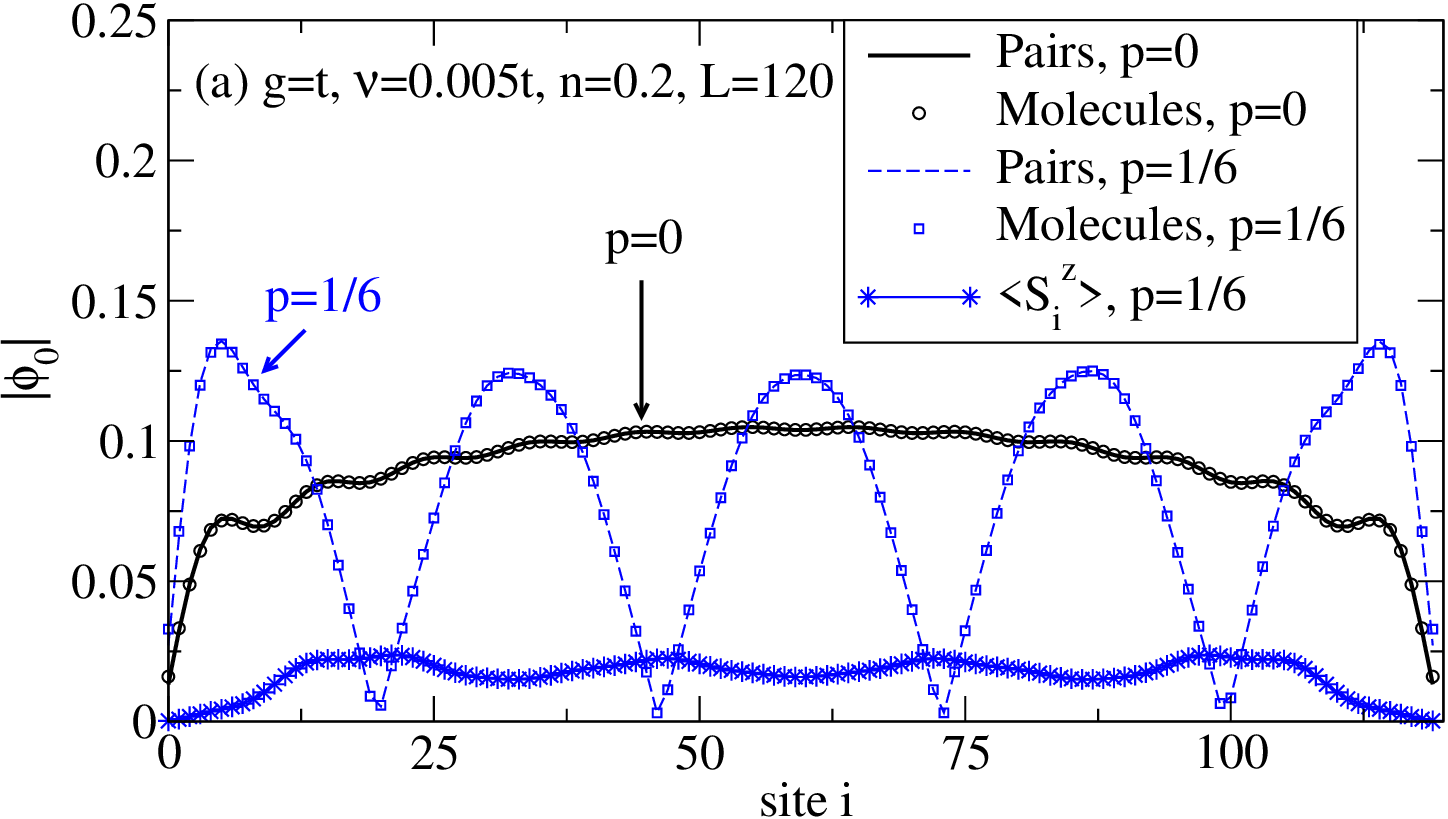}
\includegraphics[width=0.49\textwidth,angle=0]{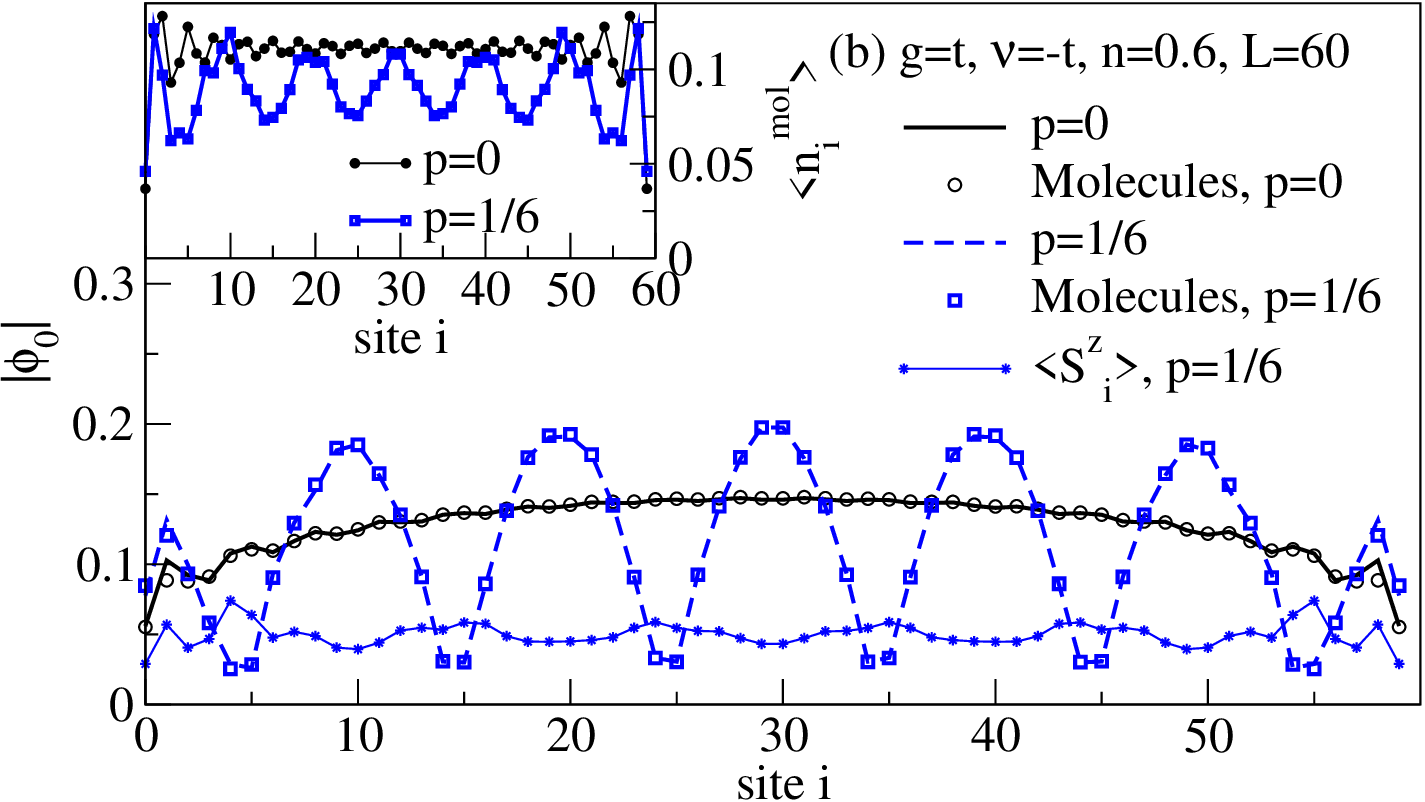}
\includegraphics[width=0.49\textwidth,angle=0]{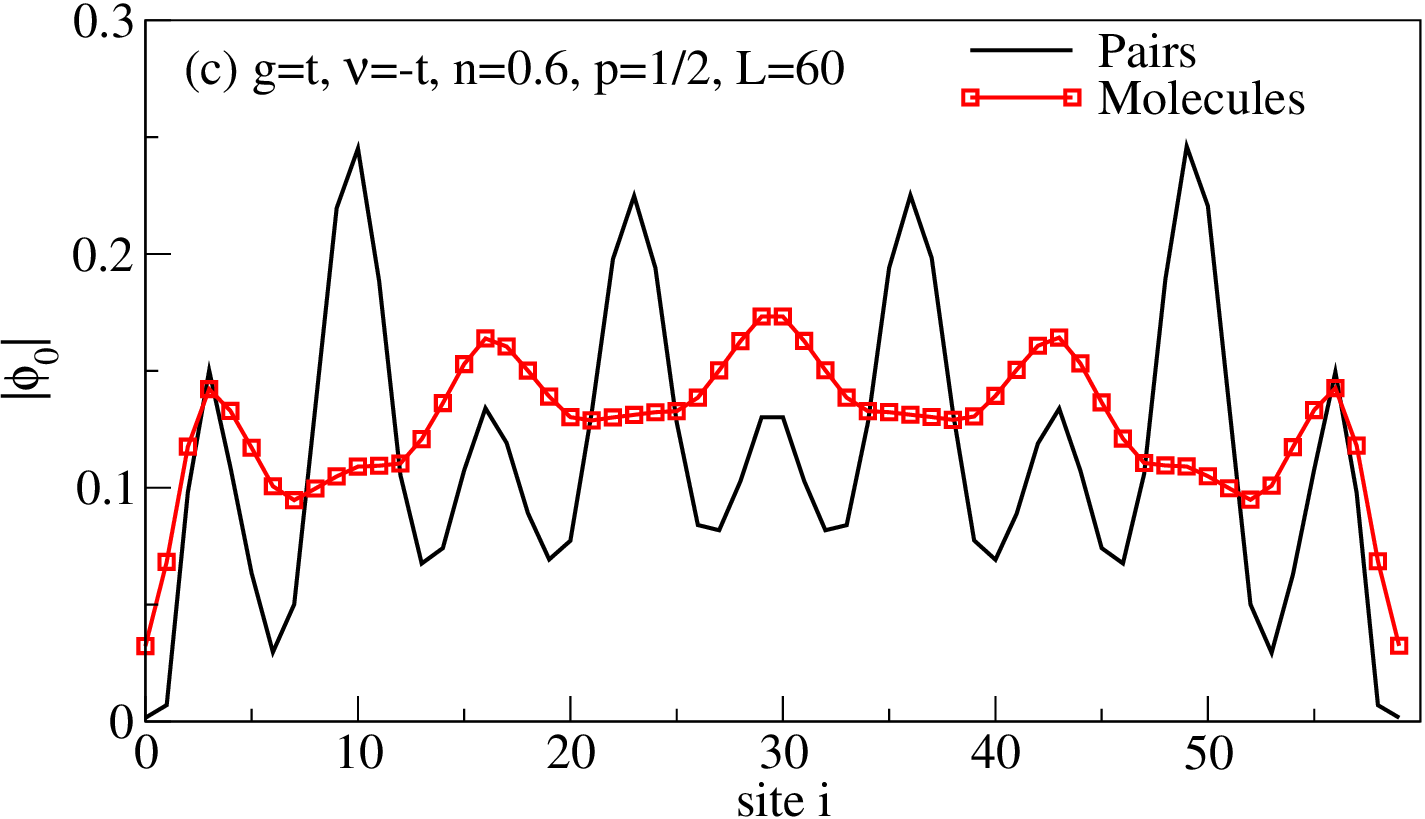}
\caption{(Color online) Natural orbital $|\phi_0|$ for 
pairs (lines) and molecules (symbols) for (a) $n=0.2$, $\nu=0.005t$  and 
(b) $n=0.6$, $\nu= -t$, both at $p=0,1/6$ (circles and squares, respectively).
The stars  are $\langle S^z_i\rangle $ for $p=1/6$. Inset in (b): density of molecules $\langle n^{\mathrm{mol}}_i\rangle $ 
at $p=0,1/6$. 
(c) Natural orbitals in the intermediate phase $p_1<p<p_2$ for $\nu=-t$, $p=1/2$, $g=t$ and $L=60$. 
} 
\label{fig:nos}
\end{figure}

To render the analysis of the locking effect \cite{recati05,citro05,orignac06,sheehy06} between $\rho_{ij}^{\mathrm{pair}}$ and $\rho_{ij}^{\mathrm{mol}}$ more  quantitative, we compute the eigenvalues and eigenvectors  of the associated one-particle density matrices, 
$\rho_{ij}^{\mathrm{pair}}$ and $\rho_{ij}^{\mathrm{mol}}$
 (the eigenvectors are sometimes called 'natural orbitals'). 
 In particular, the orbital $\phi_0$ that is connected with the largest eigenvalue
according to the Penrose-Onsager decomposition \cite{penrose56} of the density matrix 
reveals the real-space structure of the quasi-condensates \cite{footnote1}. In the presence 
of FFLO-type order,  $\phi_0$ is therefore a nontrivial function even for a homogeneous system.  
The modulus of this quantity, {\it i.e.}, $|\phi_0|$ is plotted in Fig.~\ref{fig:nos}(a) for $n=0.2$  
and in  Fig.~\ref{fig:nos}(b) for $n=0.6$; in both cases for $p=0,1/6$ and values of the detuning such that the system is in the crossover regime.

Both at $p=0$ and in the FFLO phase,   the natural orbitals of molecules and pairs are fully identical,
as has been shown for the limit of vanishing polarization in previous studies \cite{recati05,citro05}.
Further, in the {1D FFLO} phase, the spin density 
$$\langle S_i^z\rangle=(\langle n_{i,\uparrow}\rangle -\langle n_{i,\downarrow}\rangle )/2 $$ 
follows the real-space modulation of the natural orbital,
with excess majority fermions residing in the nodes of the quasi-condensate (compare Refs.~\onlinecite{feiguin07,machida84} for the case of the 
1D attractive Hubbard model).
 In contrast to the behavior of the spin density, the density of molecules  follows the modulation of the 
quasi-condensate. In other words, the molecular density has its maxima  and minima at the same positions as the natural orbital.
We should stress here that the presence of features in the densities are due to the open boundary conditions used in our simulations.
In the limit of $L\to \infty$, the density and spin profiles will become flat, while the modulations can then be detected in the
respective correlation functions (compare Refs.~\cite{rizzi08,roscilde09} for the attractive Hubbard model). 
In the experimentally relevant situation of harmonically trapped particles, however, the density profiles themselves should have 
properties similar to those discussed here for finite systems with open boundary conditions, at least in parts of the particle cloud.

Note that in the regime $p_1<p<p_2$, the molecular and the pair correlations still exhibit instabilities
at the same wave vectors [see Fig.~\ref{fig:nos}(c)], even though the  natural orbitals differ in their amplitude. The locking effect ({\it i.e.}, 
natural orbitals of pairs and molecules with the same amplitude) is re-encountered in the high-field
region $p_2<p<1$. There, the molecular $|\phi_0|$ is smooth, while the corresponding natural orbital for the pairs exhibits small oscillations.

\begin{figure}[t]
\includegraphics[width=0.49\textwidth,angle=0]{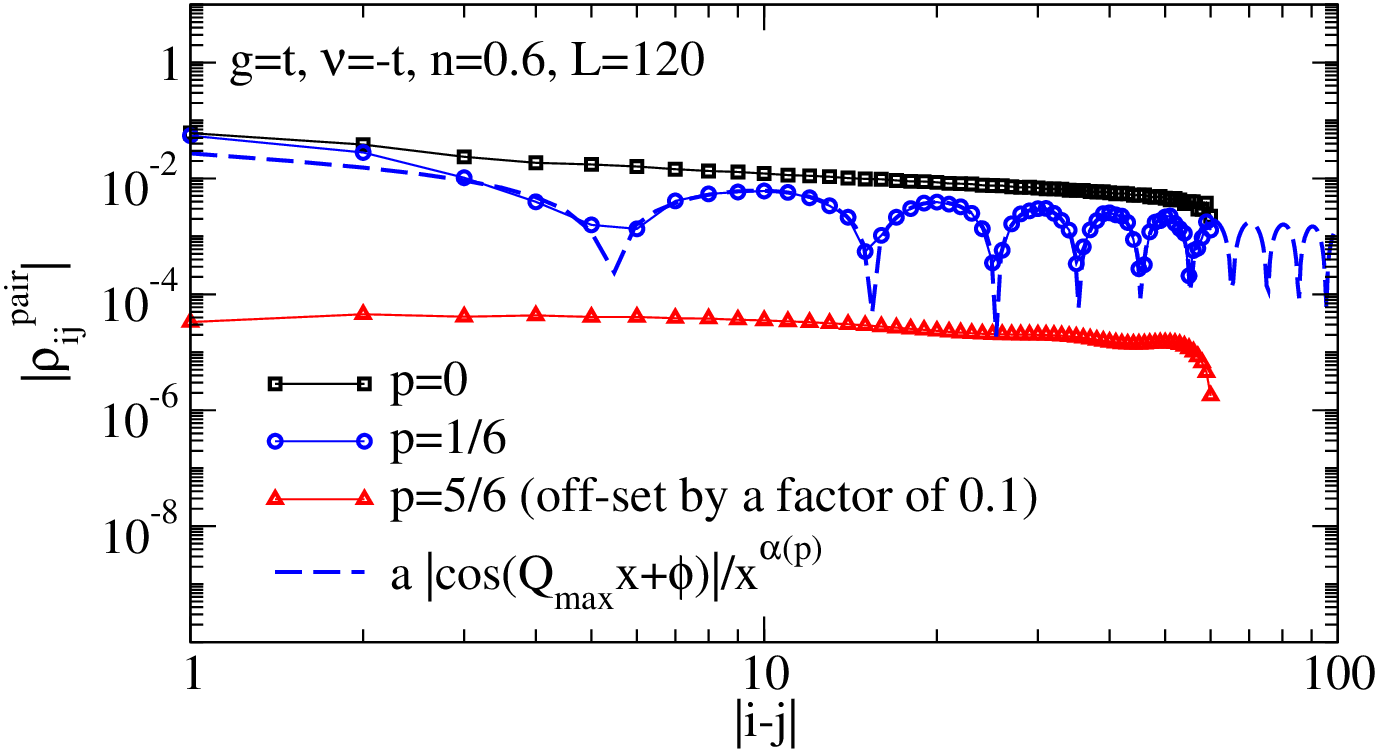}
\caption{(Color online) Decay of pair correlations in real space for $g=t$, $n=0.6$, $\nu=-t$
and $p=0,1/6,5/6$. Symbols denote DMRG results from $L=120$ sites while the
dashed line is a fit of $f(x)=a\, |\cos(Q\,x+\phi)|/x^{\alpha(p)}$ (Ref.~\onlinecite{yang01}) to the numerical data.
The fit parameters are $a,\alpha$ and $\phi$, while $Q$ is taken from the Fourier transform of the pair correlations.
The $p=5/6$ curve is off-set by a factor of 0.1 for clarity. 
} 
\label{fig:decay}
\end{figure}

\subsubsection{Spatial decay of pair correlations} 

To conclude our analysis of the pair correlations, we show that the pair correlations at $n=0.6$ 
asymptotically decay as $|\rho_{ij}^{\mathrm{pair}}| \propto 
|\cos(Q x)|/x^{\alpha}$, $x=|i-j|$, 
in agreement with predictions from bosonization for the slowest decaying contribution to $|\rho_{ij}^{\mathrm{pair}}|$ \cite{yang01}. 
To that end, we fit 
$$
f(x)=a\,|\cos(Q x+\phi)|/x^{\alpha}
$$
to our numerical data, measuring $j$ away from the center of the system ({\it i.e.}, $i=L/2$).
 Considering that the system sizes are not that large, the agreement between the DMRG results and the formula from bosonization is remarkable [see Fig.~\ref{fig:decay}].
In the regime, where  
FFLO correlations have completely disappeared, the pair correlations decay with a power law, as 
 Fig.~\ref{fig:decay} suggests for the example of $p=5/6$. Small oscillations are due to an inhomogeneous background density of pairs
 and molecules [compare the inset of Fig.¨~\ref{fig:nos}(b)].
 
 Finally, we have also verified that at $p=0$ and in the BEC limit $\nu'\gg 1$, our numerical data are consistent with a power-law decay 
 of the one-particle density matrix of the molecules
 $$
 |\rho_{ij}^{\mathrm{mol}}| \propto 1/x^{\beta}
 $$
 with an exponent of $\beta\approx 1/2 $.

\begin{figure}[h]
\includegraphics[width=0.49\textwidth,angle=0]{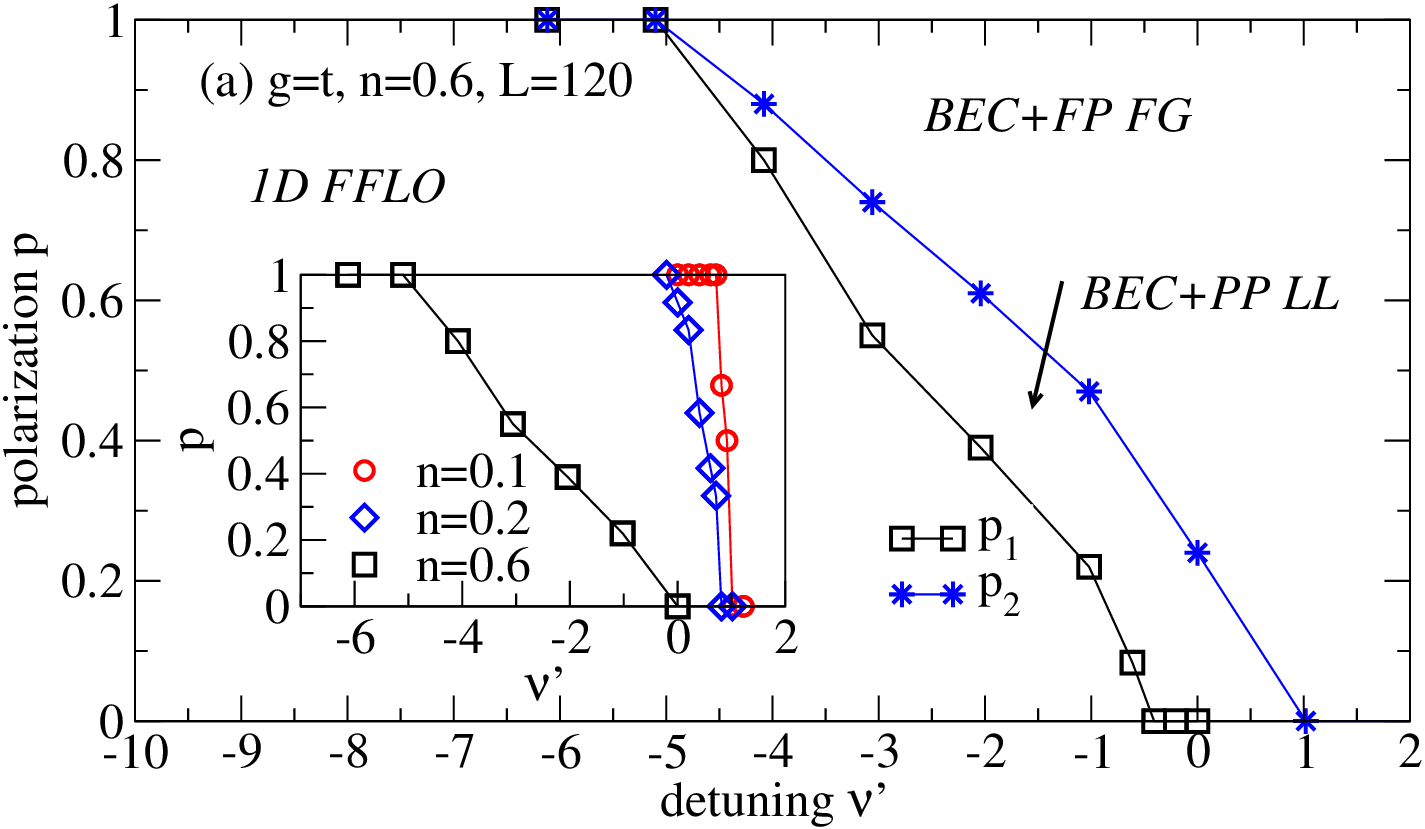}
\includegraphics[width=0.49\textwidth,angle=0]{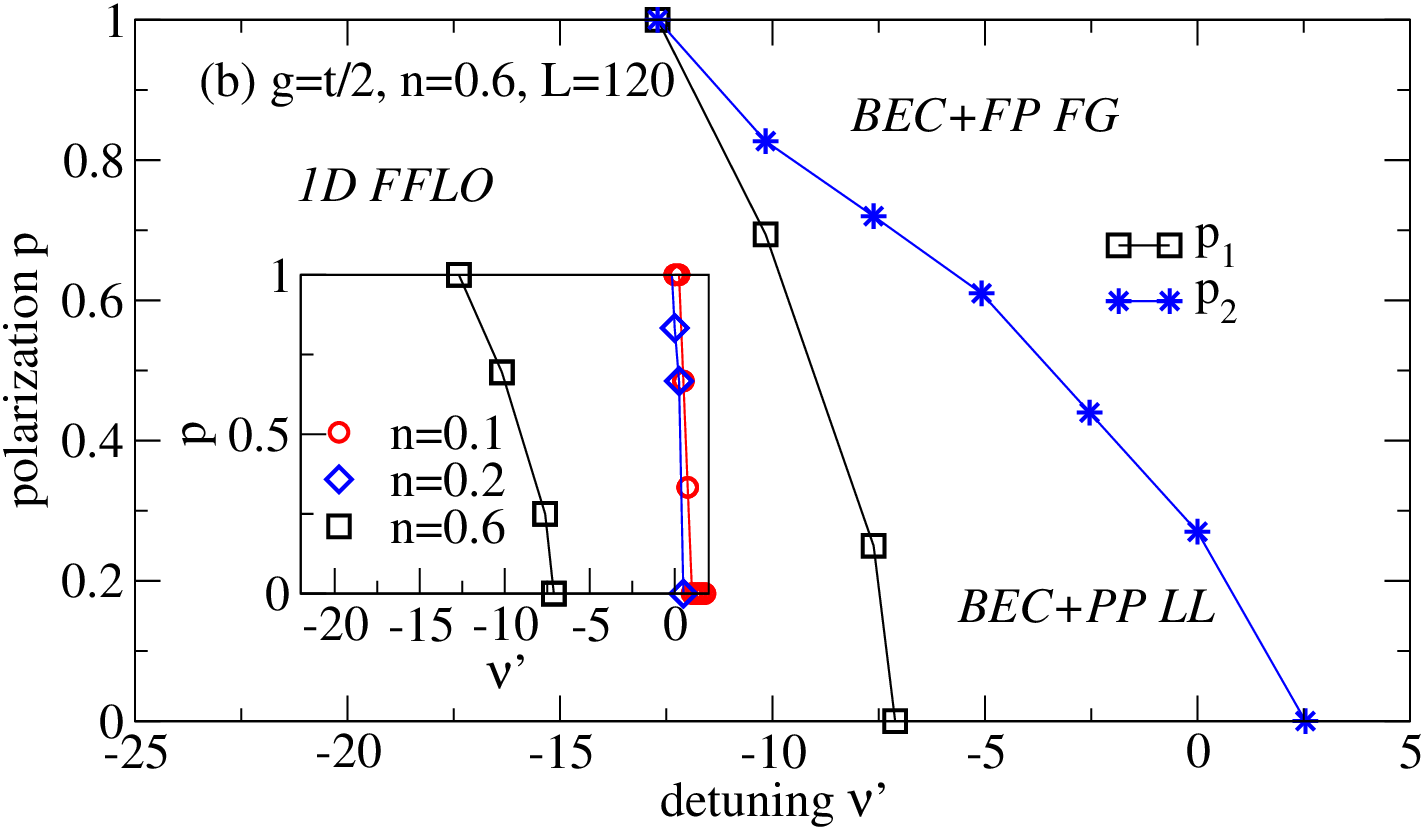}
\includegraphics[width=0.49\textwidth,angle=0]{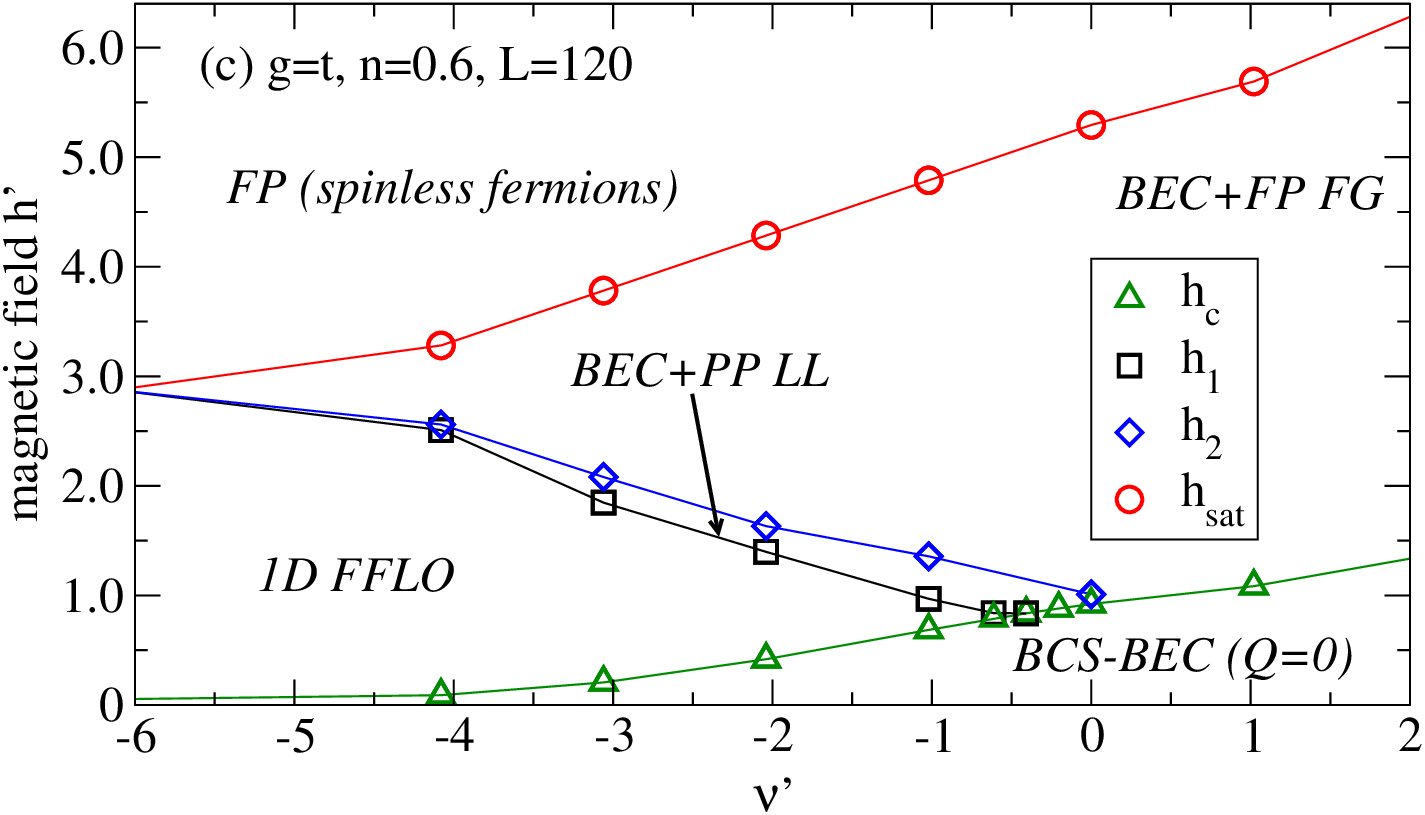}
\caption{(Color online) Phase diagrams, polarization $p$ vs. dimensionless detuning $\nu'$ for $n=0.6$
and (a) $g=t$ and (b) $g=t/2$.
The line $p=p_1(\nu')$ (squares) separates  {1D FFLO} from the {\it BEC+PP LL} regime. 
The stars  denote $p_2$ (see Sec.~\ref{sec:mol}), separating {\it BEC+PP LL}  from {\it BEC+FP FG}.
Insets in (a),(b): dependence of $p_1$  on filling: $n=0.6$ (squares), $n=0.2$ (diamonds), $n=0.1$ (circles). 
(c) The same as in (a), yet here plotted as a  magnetic field $h'$ vs. detuning $\nu'$ phase diagram. $h_c$ (triangles) is the critical field for the breakdown of the balanced gas, while
$h_{\mathrm{sat}}$ (circles) is the saturation field. Lines are guides to the eye.}
\label{fig:phase}
\end{figure}
  
\subsection{Phase diagram}
\label{sec:phase}

Our results for the phase diagram of the 1D BCS-BEC crossover described by
Eq.~\eqref{eq:ham} are summarized   in Fig.~\ref{fig:phase}, for the cases of $g=t$ [panel (a)]
and $g=t/2$ [panel (b)]. The main panels contain the data for $n=0.6$ and we present polarization $p$
vs. dimensionless $\nu'$ detuning phase diagrams.

We identify three regions at $p>0$: (i) the BEC limit, $\nu' \gg 1$ and $p_2<p<1$. Here, molecules are
 immersed into a sea of fully polarized fermions. 
 This phase is denoted as {\it BEC$+$FP FG} in the figures, 
where {\it FP FG} stands for fully polarized Fermi gas.

(ii) The {\it 1D FFLO} phase at  $0<p<p_1$. 
In the crossover regime, FFLO is suppressed as $p$ is increased. We have determined the phase boundary $p_1$ (open squares) from both the position of the first kink in the 
polarization curves and
from the pair correlations. In the latter case, at $p_1$, the peak at $Q=0$ starts to build up in the MDF of the pairs.
For instance, the {1D FFLO} phase extends up to $\nu \lesssim -0.3t$ at this filling and $g=t$. This is slightly before resonance on the BCS side, 
where, nevertheless,  the   density of molecules is already finite, {\it i.e.}, $N_{\mathrm{mol}}>0$ (compare Fig.~\ref{fig:mol}). Lastly, there is a region 
(iii) $p_1(n,\nu)<p<p_2$, beyond
which we have a $Q=0$ superfluid of molecules immersed into a  partially 
polarized ({\it PP}) fermionic gas. This third phase, denoted by {\it BEC$+$PP LL}, is eventually  replaced by the 
{\it BEC$+$FP FG} phase at $p\geq p_2$, where we determine $p_2$ from the analysis
of $p=p(h)$ curves (see Sec.~\ref{sec:mol}).

Note that the boundary of the {1D FFLO} phase,  $p_1$, depends on the filling $n$. From the insets of Figs.~\ref{fig:phase}(a) and (b), 
we infer that  the larger $n$, the wider the
crossover region is, consistent with the discussion of the number of molecules (compare Sec.~\ref{sec:nmol}). As $n\to 0$, the critical line $p=p_1$ becomes quite
steep and approaches $\nu\approx (0.048 \pm  0.002)t$, or $\nu' \approx 0.97$, for $g=t$. 

 The comparison of the $g=t$ and the $g=t/2$ phase diagram shows that the FFLO phase disappears much faster in the case of 
$g=t/2$, well before resonance. Qualitatively, one can ascribe this to the fact that 
with decreasing values of the Feshbach coupling the number of molecules, or more precisely, the closed channel fraction [compare  Eq.~\eqref{Z}] 
becomes larger. The presence of molecules
tends to reduce the number of pairs with FFLO correlations. This can be expected to more efficiently suppress FFLO physics the smaller
$g$ is since the locking of molecules and pairs is then also weaker. These observations are consistent with our DMRG results for the number of molecules and their dependence on polarization and detuning presented in Fig.~\ref{fig:mol}. In particular, the maximum number of molecules is reached at smaller values of $\nu$ the larger the polarization is.

Figure~\ref{fig:phase}(c) shows the data of panel (a) in the magnetic field vs. detuning plane, using the dimensionless detuning $\nu'$ and field, $h'=h/\epsilon^*$.
This yields additional information on the saturation field $h_{\mathrm{sat}}$ and the zero-field spin gap $\Delta$
of the standard 1D BCS-BEC crossover of the balanced system, measured by 
 $h_c$. In comparison with Fig.~\ref{fig:binding}, where we have shown
$\Delta\simeq\epsilon^*$ for $n=0.1$, we repeat that the spin gap $\Delta=2h_c$ is an increasing 
function of the filling $n$ [compare also Fig.~\ref{fig:gap}(a)]. 
In the limit of $\nu'\gg 1$,  
$\Delta= 2h_c$ behaves as $\Delta \propto \nu'$   since there, 
independently of filling, the 
ground state of the balanced system has $N\approx 2N_{\mathrm{mol}}$ and $N_f\approx 0$.

In a previous work on the three-body problem in the continuum limit,  Baur {\it et al.}~\cite{baur09} 
have shown that the change in correlations between an oscillating behavior on the BCS side
due to  FFLO physics  to a smooth one on the BEC is revealed in the symmetry of the three-body
ground state wavefunction. The numerical value of the detuning where this change occurs is  
${\nu'_c} \approx 0.63$  \cite{baur09}. It is remarkable that a similar critical value for the 
disappearance of FFLO correlations is also found in our many-body calculation of the 
phase diagram. Indeed, in the low-density limit, where a comparison makes sense, 
the boundary of the  {1D FFLO} at small polarizations is typically close to resonance, 
yet on the BEC side of positive detuning $\nu>0$.  For a quantitative comparison, 
we have determined the critical value $\nu'_c(n=0.1)$ for the loss of FFLO correlations
for several values of $g$ from data taken with $L=120$ sites and polarization $p=1/6$, 
the smallest  imbalance possible for this system size.  
The resulting values are in the range of $ 0.55\lesssim \nu'_c \lesssim 0.91$, remarkably 
close to the value inferred from three-body physics in Ref.~\cite{baur09}. 

In conclusion, it is evident from Fig.~\ref{fig:phase} that  
the best regime for observing the 1D FFLO state is (i) low density and (ii)
small polarizations. The low density will favor a large weight in the quasi-coherence peaks, while
the polarization needs to be kept smaller than $p_1$. Moreover, the {1D FFLO} phase is more stable
at large Feshbach couplings $g$.

\section{Summary and Discussion}
\label{sec:sum}

In this work, we studied the Bose-Fermi resonance model in the imbalanced case as a 
simple model to describe the BCS-BEC crossover of a spin-imbalanced system in one dimension. 
Our main focus 
 was on the existence and stability of the 1D FFLO phase.
So far, many-body calculations of 1D FFLO physics were mostly concerned with models of 
attractively interacting
fermions, which do not account for the existence of composite molecules in the closed 
channel, typically encountered in experiments. 
Using a numerically
exact method, the density matrix renormalization group method, we computed several quantities
to characterize the crossover, including the number of molecules, pair correlations, the momentum distribution function,
as well as polarization curves. Most notably, we found that FFLO correlations are suppressed 
in the crossover region due to the presence of the diatomic molecules. In particular, 
the 1D FFLO phase gives room for  a regime  of molecules, quasi-condensed at zero momentum. The latter is first immersed into  partially
polarized fermions, which is then replaced by a Bose-Fermi mixture with spinless fermions below saturation.
Thus,  the system undergoes two phase transitions in the crossover region at critical polarizations $p_1<p_2 < 1$ as the polarization increases.

While our work was concerned with the homogeneous system, in experiments, the particles
typically experience a confining harmonic potential. The shell structure 
for attractively interacting fermion models in 1D was intensely discussed. The emerging picture for the continuum
case, based on numerically or analytically exact approaches (the latter typically combined with the local density approximation)  \cite{orso07,hu07,casula08,kakashvili09}
is that one finds  either fully paired wings  at small polarization or fully polarized wings, while the core
is always partially polarized. In the case of lattice models,  DMRG calculations that   take the trap into 
account exactly  report fully polarized wings with a partially polarized core \cite{feiguin07,tezuka08} at
intermediate and large polarizations, and the latter also remains true in coupled chains at sufficiently large polarizations \cite{feiguin09}.

 While we expect the behavior of trapped, attractively interacting fermions to carry over to the
BCS regime of the Bose-Fermi resonance model, a finite density of molecules may lead to qualitatively different shell structures. 
For instance, the heavier molecules should mostly reside in the center of the trap. 
On the one hand, one may expect this to destabilize the FFLO phase in the core, while on the other hand,
as long as the Feshbach coupling $g$ and hence, the locking between pairs and molecules is sufficiently strong,
the locking could protect  the FFLO correlations.
The clarification of the effect of a harmonic trap  is left for future research.

An important question is how the FFLO state can be detected in an experiment. Several proposals
have been put forward, for instance, time-of-flight measurements \cite{yang05}, the analysis of noise correlations \cite{luescher08,paananen08}, or features
in the spin density and correlations \cite{roscilde09}. 
Regarding the spin correlations, one expects a peak at nonzero momentum $2Q\ne 0$
in the presence of FFLO order \cite{roscilde09}. In fact, the spin density follows the modulation of the natural orbitals, 
as has previously been demonstrated for the 1D attractive Hubbard model \cite{feiguin07}. As we 
showed here, 
this behavior is also realized in the FFLO phase of the Bose-Fermi-resonance model
(compare Fig.~\ref{fig:nos}). 
%While in 3D, oscillations in the density difference between majority and minority spins were experimentally observed and are often discussed as a potential
%indication for FFLO physics, 
Even if the FFLO phase was present in a 3D system,
the obstacle there is that, if at all, the FFLO phase is in the wings of a 3D, 
trapped Fermi gas (see, {\it e.g.}, Ref.~\cite{desilva06}). This constitutes another advantage 
of searching for FFLO physics in a 1D system: there, the core of a trapped gas will host this phase
\cite{orso07,feiguin07,casula08}, and therefore, the associated modulation in the 
spin density should exist in a large part of the cloud, contrary to the 3D case.

\begin{acknowledgments}
We acknowledge fruitful discussions with  A. Kolezhuk  and we thank D. Huse and the authors of Ref.~\cite{baur09}
for their comments on a previous version of this work.
U.S. and W.Z. acknowledge support from the {\it Deutsche Forschungsgemeinschaft} through FOR 801.
U.S. was further supported by the DFG through grant SCHO-621/8-2.
 F.H.-M. thanks the KITP at UCSB for its hospitality, where part of this research was carried out.
This research was supported in part by the National Science Foundation under Grant No. NSF PHY05-51164. 
We thank E. Dagotto for granting us compute time at his group's facilities at the University of Tennessee
at Knoxville.
\end{acknowledgments}

Note added in proof. Recently, we became aware of a related, very recent experiment at Rice that studies the spin-imbalanced Fermi gas with attractive interactions in one dimension \cite{liao09}.


%--------------------------------------------------------


\begin{thebibliography}{10}
\bibitem{footnote1}The absence of true long range order in one dimension
even at zero temperature implies that the largest eigenvalue of the one-particle
density matrix does not scale linearly with particle number $N$ or system size.
Instead, it behaves like $N^{1-1/(2K)}$, where $K>1$ is the associated
Luttinger parameter.
\end{thebibliography}
\end{document}